\documentclass[12pt]{iopart}
	\usepackage{textcomp}
	\usepackage{graphicx} 
	\usepackage[breaklinks=true,colorlinks,citecolor=blue,linkcolor=blue,urlcolor=blue]{hyperref}
	\usepackage{bm}
	\usepackage{color}
	\usepackage{ulem}
	\usepackage{braket}
	\usepackage{lipsum}
	\usepackage{cite}

\begin{document}

\title[Maximum power of quantum heat engines and refrigerators]{Maximum power and corresponding efficiency for two-level heat engines and refrigerators: optimality of fast cycles}

\author{P A Erdman$^1$, V Cavina$^1$, 
R Fazio$^{2,1}$, F Taddei$^3$ and V Giovannetti$^1$}

\address{$^1$ NEST, Scuola Normale Superiore and Istituto Nanoscienze-CNR, I-56126 Pisa, Italy}
\address{$^2$ ICTP, Strada Costiera 11, I-34151 Trieste, Italy}
\address{$^3$ NEST, Istituto Nanoscienze-CNR and Scuola Normale Superiore, I-56126 Pisa, Italy}
\ead{paolo.erdman@sns.it}

\vspace{10pt}
\begin{indented}
\item[]June 2019
\end{indented}

\begin{abstract}
We study how to achieve the ultimate power in the simplest, yet non trivial, model of a thermal machine, namely a two-level quantum system coupled to two thermal baths.
Without making any prior assumption on the protocol, via optimal control we show that, regardless of the microscopic details and of the operating mode of the thermal machine, the maximum power is universally achieved by a fast Otto-cycle like structure in which the controls are rapidly switched between two extremal values. A closed formula for the maximum power is derived, and finite-speed effects are discussed.
We also analyse the associated efficiency at maximum power (EMP) showing that, contrary to universal results derived in the slow-driving regime, it can approach Carnot's efficiency, no other universal bounds being allowed.
\end{abstract}

\section{Introduction}
Two thermal baths in contact through a working
fluid that can be externally driven represent the prototypical setup that has been
studied from the origin of thermodynamics up to our days.
The energy balance can be described in terms of three quantities: the work extracted from the fluid and the heat exchanged with the hot/cold baths.
The fundamental limitations to the inter-conversion of heat into work stem from the concept of irreversibility and are at the core of the second law of thermodynamics.
A working medium in contact with two baths at different temperatures is also significant from a practical point of view, since it is the paradigm behind the following specific machines: the heat engine, the refrigerator \cite{Kosloff2013,Kosloff2014,Benenti2017,Alicki2018}, the thermal accelerator \cite{Buffoni2018}, and the heater \cite{Buffoni2018}.

Quantum thermodynamics \cite{Pekola2015,Goold2016,Vinjanampathy2016} has emerged both as a field of fundamental interest, and as a potential candidate to improve the performance of thermal machines \cite{Chen1994,Rezek2006,Scully2011,Abah2012, Correa2013,Dorfman2013,Brunner2014,Zhang2014,Campisi2016,Rossnagel2016,Brandner2017,Watanabe2017,Josefsson2018,Ronzani2018,Prete2019}.
The optimal performance of these systems has been discussed within several frameworks and operational assumptions, ranging from low-dissipation and slow driving regimes\cite{Esposito2010bis,Wang2011,Ludovico2016,Cavina2017,Abiuso2018}, to shortcuts to adiabaticity approaches \cite{Deng2013,Torrontegui2013,Campo2014,Cakmak2018}, to endoreversible engines \cite{Andresen1982,Song2006}. Several techniques have been developed for the optimal control of two-level systems for achieving a variety of goals: from optimizing the speed \cite{Mukherjee2013, Bonnard2009, Zhang2015}, to generating efficient quantum gates \cite{Roloff2009, Schulte2011}, to controlling dissipation \cite{Sauer2013, Carlini2006}, and to optimizing thermodynamic performances \cite{Campisi2015,Cavina2017bis, Suri2017,Cavina2018,Pekola2019,Menczel2019}. 

\begin{figure}[!t]
	\centering
	\includegraphics[width=0.99\columnwidth]{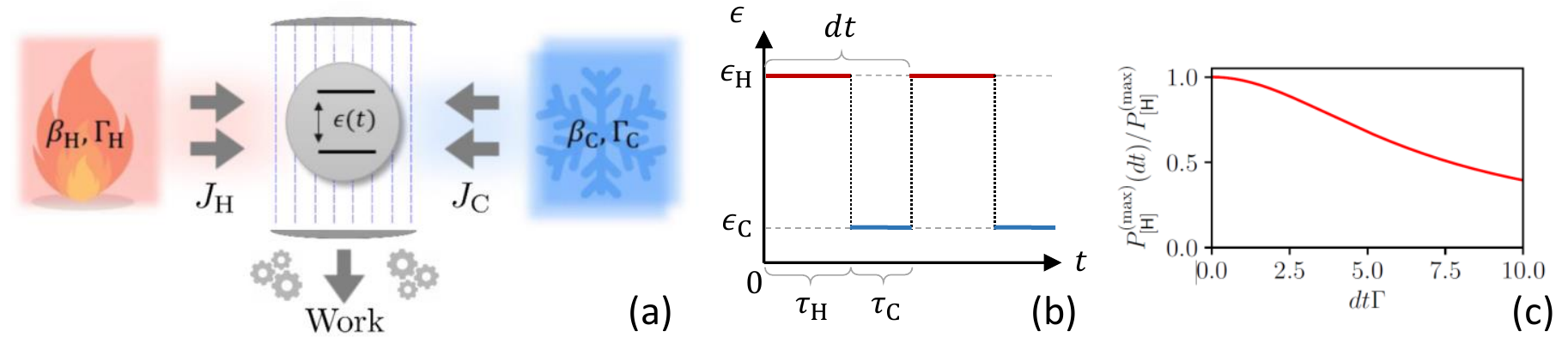}
	\caption{(a) Schematic representation of the setup. S (grey circle) is externally driven by modulating the level spacing $\epsilon(t)$ and coupled with 
	the hot bath H (red box) and the cold bath C (blue box) at inverse temperatures $\beta_\mathrm{H}$ and $\beta_\mathrm{C}$. $J_\mathrm{H}$ and $J_\mathrm{C}$ are the heat currents leaving the baths, while $\Gamma_\mathrm{H}$ and $\Gamma_\mathrm{C}$ are 
	the associated dissipation rates. Depending on the controls 
	the system can operates either as  an heat engine (mode [E]),  as a refrigerator (mode [R]), 
as a thermal accelerator (mode [A]), or as a heater (mode [H]). 
 (b) Representation of the optimal protocol that maximizes the power in the limit $dt \to 0$;  and (c) power in mode [H]  for finite values of $dt \Gamma$ normalized to the maximum power. We assume a single bath coupled to S characterized by a dissipation rate $\Gamma(\epsilon)$ such that $\Gamma(\epsilon)=\Gamma(-\epsilon)$. In this case, the maximization in Eq.~(\ref{eq:p_generic}) yields $\epsilon_\mathrm{H}^*=-\epsilon_\mathrm{C}^*$, and $\Gamma$ in (c) denotes $\Gamma(\epsilon_\mathrm{H}^*)$. }
	\label{fig:main}
\end{figure}

The aim of the present paper is to find the optimal strategy to deliver maximum power in all four previously mentioned machines. 
We perform this optimization in the simplest, yet non trivial, model of a machine which, in the spirit of quantum thermodynamics, is based on a two-level quantum system as working fluid. 
As opposed to current literature, we explicitly carry out the power maximization without making any assumptions on the operational regime, nor on the speed of the control parameters, nor on the specific coupling between the working fluid and the bath.  
We find that, if the evolution of the working medium is governed by a Markovian master equation \cite{Lindblad1976, Gorini1976}, the optimal driving takes a universal form: an infinitesimal Otto-cycle-like structure in which the control parameters must be varied between two extremal values as fast as possible. 
 This is our first main results, described in Eq.~(\ref{eq:p_generic}). Surprisingly, the optimal solution is achieved in the ``fast-driving'' regime, i.e. when the driving frequency is faster than the typical dissipation rate induced by the baths, which has received little attention in literature \cite{Feldmann1996,Feldmann2000,Cerino2016}. 

By applying our optimal protocol to heat engines and refrigerators, we find new theoretical bounds on the efficiency at maximum power (EMP). Many upper limits to the EMP, strictly smaller than Carnot's efficiency, have been derived in literature, such as the Curzon-Ahlborn and Schmiedl-Seifert efficiencies. The Curzon-Ahlborn efficiency emerges in various specific models \cite{Curzon1975, Chambadal1957, Novikov1957}, and it has been derived by general arguments from linear irreversible thermodynamics \cite{Broeck2005}. The Schmiedl-Seifert efficiency has been proven to be universal in cyclic Brownian heat engines \cite{Schmiedl2007} and for any driven system operating in the slow-driving regime \cite{Esposito2010bis}. By studying the efficiency of our system at the ultimate power, i.e. in the fast-driving regime, we prove that there is no fundamental upper bound to the EMP. Indeed, we show that the Carnot efficiency is reachable \textit{at maximum power} through a suitable engineering of the bath couplings.
 This is our second main results, illustrated in Figs.~\ref{fig:eta_max_pow}b, \ref{fig:eta_max_pow}c and \ref{fig:cop_max_pow}. 
In view of experimental implementations, we assess the impact of finite-time effects on our optimal protocol,  finding that the maximum power does not decrease much if the external driving is not much slower than the typical dissipation rate induced by the baths \cite{Koski2014,Maillet2018}.
 Furthermore, we apply our optimal protocol to two experimentally accessible models, namely photonic baths coupled to a qubit \cite{Breuer2002, Geva1992, Alicki1979, Ronzani2018, Senior2019} and electronic leads coupled to a quantum dot \cite{Esposito2009,Esposito2010,Koski2014,Maillet2018,Josefsson2018, Prete2019}.

\section{Maximum Power.} 
The setup we consider  consists of a two-level quantum system S 
with energy gap $\epsilon(t)$ that can be externally modulated \cite{note1}.
As schematically shown in 
Fig.~\ref{fig:main}a, the system is placed in thermal contact with two reservoirs, the hot  bath H at inverse temperature $\beta_{\rm H}$  and the cold bath C at inverse temperature $\beta_{\rm C}$,
respectively characterized by coupling constants $\lambda_\mathrm{H}(t)$ and $\lambda_\mathrm{C}(t)$ that can be
modulated in time.
The system can operate in four different modes: 
  i) the heat engine mode [E], where  S is used to produce work 
by extracting heat from H while donating it to C;
ii)  the refrigerator mode  [R], where S is used to extract heat from C;
iii) the thermal accelerator mode [A], where S operates to move as much heat as possible to C; 
iv)  the heater mode [H], where we simply use S to deliver as much heat as possible to both H and C.
Assuming cyclic modulation of the controls (i.e. of $\epsilon(t)$, $\lambda_\mathrm{H}(t)$ and $\lambda_\mathrm{C}(t)$) we are interested in maximizing the corresponding 
averaged output powers of each operating mode, i.e. the quantities 
 \begin{eqnarray}
	&P_{\mathrm{[E]}} = \braket{J_{\mathrm{H}}}+\braket{{J}_\mathrm{C}},  \qquad
	&P_{\mathrm{[R]}} = \braket{J_\mathrm{C}}, \\
	&P_{\mathrm{[A]}} =- \braket{J_\mathrm{C}},
	&P_{\mathrm{[H]}} = -\braket{J_\mathrm{H}}-\braket{J_\mathrm{C}},
\label{eq:max_pow_def}
\end{eqnarray}
 where   $J_\mathrm{H}$ and $J_\mathrm{C}$ are the instantaneous heat fluxes
entering the hot and cold reservoirs respectively, and where 
the symbol $\braket{\cdots}$ stands for temporal average over a modulation cycle of the controls. 
To tackle the problem 
we adopt 
a  Markovian Master Equation (MME)  approach~\cite{Breuer2002}, namely we write
\begin{equation}
\frac{d}{dt} \hat{\rho}  =  -\frac{i}{\hbar}[\hat{\mathcal{H}}, \hat{\rho}]_- + \sum_{\alpha={\rm H},{\rm C}} \mathcal{D}_{\alpha}\left[\hat{\rho} \right],   
\end{equation}
where $\hat{\rho}$ is the density matrix of the two-level system at time $t$,  
 $\hat{\mathcal{H}} := \epsilon(t)\hat{\sigma}_+\hat{\sigma}_-$  
 its  local Hamiltonian, and  
 \begin{equation}
 \mathcal{D}_\alpha\left[\cdots\right] := \sum_{i=\pm} \lambda_\alpha(t)\Gamma^{(i)}_\alpha(\epsilon(t))( \hat{\sigma}_i \cdots  \hat{\sigma}_i^\dagger -\frac{1}{2} [ \hat{\sigma}_i^\dagger\hat{\sigma}_i, \cdots ]_+)
 \end{equation}
is the Gorini-Kossakowski-Sudarshan-Lindblad  dissipator~\cite{Gorini1976,Lindblad1976} associated with the bath $\alpha={\rm H}, {\rm C}$.
We have denoted with $\hat{\sigma}_+$ and $\hat{\sigma}_-$  the raising and lowering operators of S and 
with the symbol $[\cdots,\cdots]_{\mp}$ the commutator ($-$)  and anti-commutator ($+$) operations.
$\cal{D}_{\alpha}$ is
 characterized by dissipation rates $\Gamma_{\alpha}^{(i=\pm)}(\epsilon)$ 
 and by the dimensionless  coupling constant $\lambda_{\alpha}(t)\in [0,1]$ that plays the
 role of a ``switch'' control parameter.
 It is worth noticing that, since $[\hat{\mathcal{H}}(t), \hat{\mathcal{H}}(t')]=0$, the MME we employ is valid also in the fast-driving regime, provided that the correlation time of the bath is the smallest timescale in our problem \cite{dann2018}. Therefore, the fast-driving regime is characterized by a control frequency which is faster than the typical dissipation rate, but slower than the inverse correlation time of the bath. Furthermore, we assume that the Hamiltonain $\hat{\mathcal{H}}_{\mathrm{int}}$, describing the system-bath interaction, is such that its expectation value on the Gibbs state of the baths is zero (this is true, for example, for tunnel-like Hamiltonians, where the number of creation/annihilation operators of the bath entering $\hat{\mathcal{H}}_{\mathrm{int}}$ is odd). Such assumption guarantees that no work is necessary to switch on and off the coupling between the system and the baths.

Without assigning any specific value to the dissipation rates, we only require them to
obey the detailed balance equation 
	${\Gamma^{(+)}_\alpha(\epsilon )}/{\Gamma^{(-)}_\alpha(\epsilon)} = e^{-\beta_\alpha  \epsilon}$.
This ensures that,
 at constant level spacing $\epsilon$, the system S will  evolve into a thermal Gibbs state characterized by 
an excitation probability 
\begin{equation}
	p_{\mathrm{eq}}^{(\alpha)}(\epsilon) := \frac{\Gamma^{(+)}_\alpha(\epsilon)}{\Gamma^{(+)}_\alpha(\epsilon) + \Gamma^{(-)}_\alpha(\epsilon)  }= \frac{1}{1+e^{\beta_\alpha\epsilon}}
\end{equation}
  when in contact only with heat bath $\alpha$.
For simplicity, we consider the system to be coupled to one heat bath at the time, i.e. we assume that $\lambda_\mathrm{H}(t)+\lambda_\mathrm{C}(t)=1$, and that $\lambda_\alpha(t)$ can take the values $0$ or $1$.
  As we shall see in the following, this constraint, as well as the possibility of controlling the coupling constants $\lambda_\alpha(t)$, is not fundamental 
 to derive our results, at least for those cases where the effective dissipation rate
   \begin{eqnarray} \Gamma_\alpha(\epsilon) := \Gamma^{(+)}_\alpha(\epsilon)+\Gamma^{(-)}_\alpha(\epsilon)\;
\label{GAMMA}\end{eqnarray}
of each bath is sufficiently peaked around distinct values.
The instantaneous heat flux leaving the thermal bath $\alpha$ 
can now be expressed as \cite{Alicki1979} 
\begin{equation*}
	J_\alpha = \mathrm{tr}[ \hat{\mathcal{H}} \mathcal{D}_\alpha\left[\hat{\rho} \right] ] = 
	- \epsilon(t) \lambda_\alpha(t) \Gamma_\alpha[\epsilon(t)] (p(t)-
		p_{\mathrm{eq}}^{(\alpha)}[\epsilon(t)]),
\end{equation*}
  where $p(t) := \mathrm{tr}[\hat{\sigma}_+\hat{\sigma}_-\hat{\rho}(t)]$  is the probability of finding S in the excited state of ${\cal H}$  at time $t$ which 
  obeys the following differential equation 
\begin{eqnarray} 
\frac{d}{dt}p(t)  = - \sum_{\alpha={\rm H,C}}  \lambda_\alpha(t)\Gamma_\alpha[\epsilon(t)] (p(t)-p_{\mathrm{eq}}^{(\alpha)}[\epsilon(t)]),	\label{eq:lindblad1}
\end{eqnarray}  
according to the MME specified above.
 By explicit integration of~(\ref{eq:lindblad1}) we can hence  transform all the terms in Eq.~(\ref{eq:max_pow_def}) into functionals of the controls
 which can then be optimized with respect to all possible choices of the latter. 
 
 As shown in App.~A, we find that the protocols which maximize the average power of a fixed physical setup, i.e. at fixed dissipation rates, are cycles performed in the fast-driving regime, i.e. when the driving frequency is faster than the typical dissipation rate. More precisely, the optimal cycle is such that $\epsilon(t)$ instantaneously jumps between two values $\epsilon_\mathrm{H}$ and $\epsilon_\mathrm{C}$, see Fig.~\ref{fig:main}b, while being in contact, respectively, only with the hot and cold bath for infinitesimal times $\tau_\mathrm{H}$ and $\tau_\mathrm{C}$ fulfilling  the condition 
	${\tau_\mathrm{H}}/{\tau_\mathrm{C}} = \sqrt{{\Gamma_\mathrm{C}(\epsilon_{\rm C})}/{\Gamma_\mathrm{H}(\epsilon_{\rm H})}}$ \cite{NOTA}.
As in Otto cycles considered in literature  (see the
extensive literature on this topic, e.g. \cite{Rezek2006,Quan2007,Abah2012,Karimi2016,Kosloff2017,Watanabe2017}), no heat is transferred during the jumps and no work is done while the system is in contact with the baths.
 The resulting maximum power averaged over one period   can then be cast into the following compact expression (see App. B for details)
\begin{equation}
	P^{(\max)}_{[\nu]} = \max_{(\epsilon_\mathrm{H}, \epsilon_\mathrm{C})
	 \in \mathcal{C}}  \frac{\Gamma_\mathrm{H}(\epsilon_{\rm H})\Gamma_\mathrm{C}(\epsilon_{\rm C})   \left( p_{\mathrm{eq}}^{({\rm H})}(\epsilon_\mathrm{H}) - p_{\mathrm{eq}}^{({\rm C})}(\epsilon_\mathrm{C}) \right)}{ \left(\sqrt{\Gamma_\mathrm{H}(\epsilon_{\rm H})} +\sqrt{\Gamma_\mathrm{C}(\epsilon_{\rm C})} \right) ^2}\; \widetilde{\epsilon}_{[\nu]}  ,
	\label{eq:p_generic}
\end{equation} 
 where $\nu=$ E,R,A,H and the quantity 
 $ \widetilde{\epsilon}_{[\nu]}$ is given 
 by 
$\widetilde{\epsilon}_{[E]} = \epsilon_\mathrm{H}  -\epsilon_\mathrm{C}$, $\widetilde{\epsilon}_{\mathrm{[R]}} = -\epsilon_\mathrm{C}$, $\widetilde{\epsilon}_{\mathrm{[A]}} = \epsilon_\mathrm{C}$, and $\widetilde{\epsilon}_{[H]} = \epsilon_\mathrm{C}  -\epsilon_\mathrm{H}$.
In Eq. (\ref{eq:p_generic}) $\mathcal{C}$ is the range over which the energy gap $\epsilon(t)$ of S is allowed to be varied according to the possible technical limitations associated with the specific
implementation of the setup. 

Equation~(\ref{eq:p_generic}), which stems from the optimality of the fast-driving regime, is the first main result of the present work. We emphasize that, as opposed to current literature, our closed expression for the maximum power holds for \textit{any} dissipation rate function $\Gamma_\mathrm{H/C}(\epsilon)$.
In the following we will apply our result to specific implementations
 which are relevant experimentally and compute their associated efficiencies at maximum power.
In particular we shall consider 
 the 
 case of fermionic ($\mathrm{F}_n$) and bosonic ($\mathrm{B}_n$) baths
 with associated effective rates of the form 
\begin{eqnarray}
    &\Gamma^{(\mathrm{F}_n)}_\alpha(\epsilon) = k_\alpha \epsilon^n, 
    &\Gamma^{(\mathrm{B}_n)}_\alpha(\epsilon) = k_\alpha \epsilon^n \coth{\left(\beta_\alpha\epsilon/2\right)},
    \label{eq:exp_rates}
\end{eqnarray}
with $n\geq 0$ integer and with $k_\alpha$ being a coupling strength constant.
The fermionic rate (the first of Eq. (\ref{eq:exp_rates})) 
for instance can describe two electronic leads, with density of states depending on $n$, tunnel coupled to a single-level quantum-dot \cite{Beenakker1991, Esposito2010, Erdman2017};
the bosonic one instead is applied in the study
of two-level atoms in a dispersive quantum electromagnetic 
cavity \cite{Wang2012pre}.

\section{Heat engine mode $[E]$.}
It is common belief that  the efficiency of a heat engine (work extracted over heat absorbed from the hot bath H), driven at maximum power (EMP), should exhibit
 a finite gap with respect to the Carnot
efficiency $\eta_\mathrm{c} :=1-\beta_\mathrm{H}/\beta_\mathrm{C}$. Indeed, this is corroborated by various results on EMP bounds: the Curzon-Ahlborn EMP  $\eta_{\mathrm{CA}} := 1- \sqrt{1-\eta_\mathrm{c}}$ ~ emerges in various specific models\cite{Curzon1975,Schmiedl2007,Cavina2017}, and it has been derived by general arguments from linear irreversible thermodynamics \cite{Broeck2005}, while the Schmiedl-Seifert EMP  $\eta_{\mathrm{SS}} := \eta_\mathrm{c}/(2-\eta_\mathrm{c})$ has been proven to be universal for any driven system operating in the slow-driving regime \cite{Esposito2010bis}. However, the completely out-of-equilibrium and optimal cycles associated with the values of $P^{(\max)}_{[E]}$ reported in  Eq.~(\ref{eq:p_generic}), do not fulfill such assumptions.
As a matter of fact, by choosing particular  ``energy filtering'' dissipation rates $\Gamma_\alpha(\epsilon)$ (instead of the regular ones given e.g. in Eq.~(\ref{eq:exp_rates})), we can produce configurations which approach Carnot's efficiency with arbitrary precision while delivering maximum power, proving the lack of any fundamental bound to the EMP.
Before discussing this highly not trivial effect, it is worth analyzing the performances associated with the
baths models of Eq.~(\ref{eq:exp_rates}). 

We remind that the efficiency 
 of an Otto cycle heat engine 
  working between the internal energies $\epsilon_\mathrm{C}$ and $\epsilon_\mathrm{H}$ is given by
  $\eta = 1 - {\epsilon_\mathrm{C}}/{\epsilon_\mathrm{H}}$. 
Accordingly, indicating with
$\epsilon_\mathrm{H}^*$ and $\epsilon_\mathrm{C}^*$ the values of the gaps that lead to 
the maximum of the r.h.s term of Eq.~(\ref{eq:p_generic}), 
 we write
 the  EMP  of our scheme as 
 \begin{eqnarray}\label{EMP1}  \eta(P^{(\max)}_{[\mathrm{E}]}) = 1 - {\epsilon^*_\mathrm{C}}/{\epsilon^*_\mathrm{H}}=
  1 - \left(1 - \eta_\mathrm{c} \right)	{\epsilon^*_\mathrm{C}\beta_\mathrm{C}}/{\epsilon^*_\mathrm{H}\beta_\mathrm{H}}\;.\end{eqnarray}  
  \begin{figure}[!tb]
	\centering
	\includegraphics[width=0.49\columnwidth]{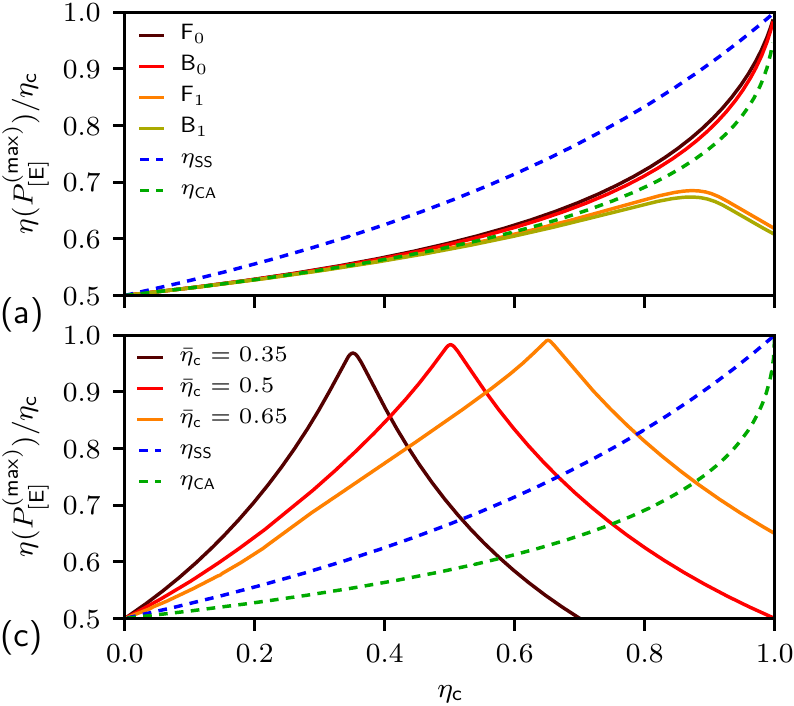}
	\includegraphics[width=0.49\columnwidth]{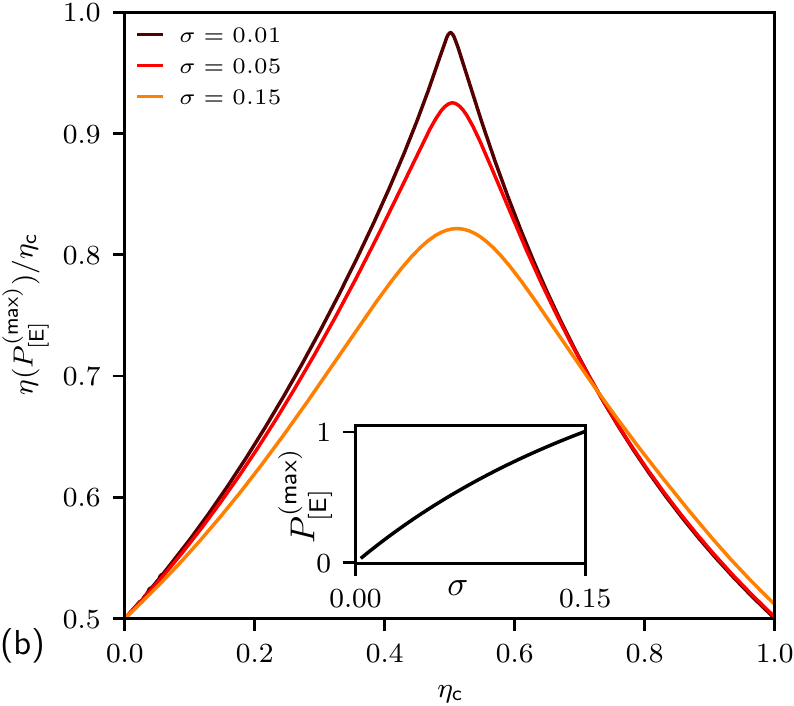}
	\caption{EMP for the heat engine mode [$\eta(P^{(\max)}_{[\mathrm{E}]})$ of Eq.~(\ref{EMP1})], normalized to $\eta_\mathrm{c}$, as a function of $\eta_\mathrm{c}$ (varied by fixing $\beta_\mathrm{H}$ and sweeping over $\beta_\mathrm{C}$). (a) shows $\eta(P^{(\max)}_{[\mathrm{E}]})$ for the fermionic models ($\mathrm{F}_0$ and $\mathrm{F}_1$) and the bosonic models ($\mathrm{B}_0$ and $\mathrm{B}_1$) of Eq.~(\ref{eq:exp_rates})
	 together with the upper bounds $\eta_{\mathrm{SS}}$ \cite{Schmiedl2007} and $\eta_{\mathrm{CA}}$ \cite{Curzon1975}.
	 Notice that as $\eta_{\mathrm{c}}\rightarrow 0$ (small baths
	temperature difference), we have $\eta(P^{(\max)}_{[\mathrm{E}]})\simeq \eta_{\mathrm{c}}/2 + \eta_\mathrm{c}^2/8$ as expected. 
	For $\eta_{\mathrm{c}}\to 1$, instead,  the value of $\eta(P^{\max}_{[\mathrm{E}]})$ for the models $\mathrm{F}_1$ and $\mathrm{B}_1$ saturates to a finite fraction of $\eta_{\mathrm{c}}$, while the $\mathrm{F}_0$ and $\mathrm{B}_0$ models reach Carnot efficiency. 
	The Fermionic model displays a slightly larger $\eta(P^{(\max)}_{[\mathrm{E}]})$ than the corresponding bosonic model. 
	In all models we consider
  symmetric leads, i.e. $k_\mathrm{H}=k_\mathrm{C}$. Note that $\eta(P^{(\max)}_{[\mathrm{E}]})$ does not depend on the value of $k_\alpha$.
  (b) and (c) show $\eta(P^{(\max)}_{[\mathrm{E}]})$ computed using Lorentzian filtering rates $\Gamma_\alpha(\epsilon_\alpha) = \gamma \sigma^2 /(\sigma^2+({\epsilon_\alpha-\bar{\epsilon}_\alpha})^2)$ with $\gamma$, $\sigma$ and $\bar{\epsilon}_\alpha$ positive constants (systems with multiple quantum-dots in series \cite{Wegewijs1999} e.g. exhibit such dependence).
	  In both panels we fix $\bar{\epsilon}_\mathrm{C}= 1$.
	  (b): we set $\bar{\epsilon}_\mathrm{H} = 2 \bar{\epsilon}_\mathrm{C}$ such that we expect to approach $\eta_\mathrm{c}$ at $\bar{\eta}_\mathrm{c} = 1/2$. 
	 Indeed, as $\sigma$ decreases, $\eta(P^{(\max)}_{[\mathrm{E}]})/\eta_\mathrm{c}$ approaches one at $\bar{\eta}_\mathrm{c}=1/2$. Conversely, the corresponding maximum power decreases: in the inset, where $P^{(\max)}_{[\mathrm{E}]}$ is plotted as a function of $\sigma$ for $\bar{\eta}_\mathrm{c}=1/2$, we see that the maximum power becomes vanishingly small for $\sigma\to 0$. The power is normalized to the its value for $\sigma=0.15$, where $P^{(\max)}_{[\mathrm{E}]}=0.0044 \, \gamma\beta_\mathrm{H}^{-1}$.
	 (c): at fixed $\sigma=0.01$, we show that the EMP can approach $\eta_\mathrm{c}$ at any bath temperature. We choose $\bar{\epsilon}_\mathrm{H}/\bar{\epsilon}_\mathrm{C} =$ $1/0.65$, $1/0.5$ and $1/0.35$, corresponding to $\bar{\eta}_\mathrm{c} =$ $0.35$, $0.5$ and $0.65$. Energies are expressed in units of $1/\beta_\mathrm{H}$, and the EMP does not depend on $\gamma$.
  }
	\label{fig:eta_max_pow}
\end{figure}
In Fig.~\ref{fig:eta_max_pow}a we report  the value of $\eta(P^{(\max)}_{[\mathrm{E}]})$ obtained
from~(\ref{EMP1})  for 
  the  rates of Eq.~(\ref{eq:exp_rates}) for $n=0,1$.
 By a direct comparison with $\eta_{\mathrm{CA}}$ and $\eta_{\mathrm{SS}}$,
 one notices that while the second is always respected by our optimal protocol, the first is
 outperformed at least for the baths $\mathrm{F}_0$ and $\mathrm{B}_0$, confirming the findings of 
 Refs.~\cite{Esposito2009, Erdman2017, Cavina2018}. For small temperature differences between the baths,  
the EMP can be expanded as
a power series in  $\eta_\mathrm{c}$ of the form 
	 $a_1\eta_\mathrm{c} + a_2\eta_\mathrm{c}^2 + \cdots.$
It has been shown that $a_1=1/2$ is a universal property of low dissipation heat engines \cite{Esposito2010bis} and, in this context, $a_2=1/8$ is associated with symmetric dissipation coefficients. As explicitly discussed in App. C, we find that also our protocol delivers an efficiency at maximum power with
a first order expansion term $a_1=1/2$ and with a second order correction $a_2=1/8$
achieved if we assume that the two leads are symmetric, i.e. $\Gamma_\mathrm{H}(\epsilon,\beta) = \Gamma_\mathrm{C}(\epsilon,\beta)$, or if the rates are constants. 

We now turn to the possibility of having $\eta(P^{(\max)}_{[\mathrm{E}]})$ arbitrarily close to $\eta_{\mathrm{c}}$.
By a close inspection of the second identity of Eq.~(\ref{EMP1}) we notice that one can have
$\eta(P^{(\max)}_{[\mathrm{E}]})\simeq \eta_{\mathrm{c}}$
for all those models where
the maximum power (see Eq.~(\ref{eq:p_generic})) is obtained for values of the gaps fulfilling the condition 
$\epsilon^*_\mathrm{C}\beta_\mathrm{C}\approx\epsilon^*_\mathrm{H}\beta_\mathrm{H}$.
Consider hence  a scenario where the rates $\Gamma_\alpha(\epsilon_\alpha)$ are
such that the power is vanishingly small for all values of $\epsilon_\alpha$ except for a windows of width $\sigma$ around a given value $\bar{\epsilon}_\alpha$, a configuration that can be used to eliminate the presence of the activation controls $\lambda_\alpha(t)$ from the problem. 
Under the assumption of small enough $\sigma$, we expect the maximization in Eq.~(\ref{eq:p_generic}) to yield $\beta_\mathrm{C}\epsilon^*_\mathrm{C} \approx \beta_\mathrm{H}\epsilon^*_\mathrm{H}$ when the inverse temperature ratio is $\beta_\mathrm{C}/\beta_\mathrm{H} \approx \bar{\epsilon}_\mathrm{H}/\bar{\epsilon}_\mathrm{C}$, so that $\eta(P^{(\max)}_{[\mathrm{E}]}) \approx \eta_\mathrm{c}$.
This is indeed evident from Fig.~\ref{fig:eta_max_pow}b and \ref{fig:eta_max_pow}c, where we report the value  $\eta(P^{(\max)}_{[\mathrm{E}]})$
as a function of $\eta_\mathrm{c}$ (which represents the temperature of the baths) for rates having a Lorentzian shape dependence: by decreasing  $\sigma$, 
the EMP approaches Carnot's efficiency at $\bar{\eta}_\mathrm{c}:=1-\bar{\epsilon}_\mathrm{C}/\bar{\epsilon}_\mathrm{H}=1/2$ (Fig. \ref{fig:eta_max_pow}b), while by tuning the position of the peak of the Lorentzian rates, the EMP can approach Carnot's efficiency at any given bath temperature configuration $\bar{\eta}_\mathrm{c}$ (Fig. \ref{fig:eta_max_pow}c). We emphasize that even our system with Lorentzian shaped rates would exhibit an EMP bounded by $\eta_{\mathrm{SS}}$ if operated in the slow-driving regime. The possibility of reaching Carnot's efficiency at maximum power is thus a characteristic which emerges thanks to the fast-driving regime. Conversely, as $\sigma$ decreases and $\eta(P^{(\max)}_{[\mathrm{E}]}) \to \eta_\mathrm{c}$, the corresponding maximum power tends to zero (see the inset of Fig.~\ref{fig:eta_max_pow}b where the maximum power, at $\bar{\eta}_\mathrm{c}=1/2$, is plotted as a function of $\sigma$).

\section{Refrigerator mode $[{\rm R}]$.}
 The efficiency of a refrigerator is quantified by the coefficient of performance (COP), i.e.
the ratio between the heat extracted from the cold bath and the work done on the system. For an Otto-cylce the COP is given by  $\mathrm{C}_{\mathrm{op}}= {\epsilon_\mathrm{C}}/({\epsilon_\mathrm{H}-\epsilon_\mathrm{C}})$
which, by replacing the values $\epsilon^*_\mathrm{C}$, $\epsilon^*_\mathrm{H}$ that lead to the
maximum  $P^{(\max)}_{[R]}$ of Eq.~(\ref{eq:p_generic}), 
yields an associated COP at maximum power (CMP)
 equal to 
 \begin{eqnarray} 
 	\mathrm{C}_{\mathrm{op}}(P^{\max}_{[\mathrm{R}]}) = \frac{\epsilon^*_\mathrm{C}}{({\epsilon^*_\mathrm{H}-\epsilon^*_\mathrm{C}})} = \left[ \frac{\beta_\mathrm{H}\epsilon^*_\mathrm{H}}{\beta_\mathrm{C}\epsilon^*_\mathrm{C}}(1/\mathrm{C}_{\mathrm{op}}^{(\mathrm{c})} + 1) -1 \right]^{-1}
	\label{QUESTAQUI} 
\end{eqnarray} 
where 
 $\mathrm{C}_{\mathrm{op}}^{(\mathrm{c})}  := \beta_\mathrm{C}^{-1}/(\beta_\mathrm{H}^{-1}-\beta_\mathrm{C}^{-1})$ is the maximum COP dictated by the second law.
Remarkably, as in the heat engine case, we can produce configurations which approach $\mathrm{C}_{\mathrm{op}}^{(\mathrm{c})}$  with arbitrary precision while delivering maximum power exploiting the same ``energy filtering'' dissipation rates. Before discussing this effect we present some universal properties of the CMP and we analyze the performance of the baths models of Eq.~(\ref{eq:exp_rates}).
 
Assuming that the rates depend on the energy and on the temperature through the product $\beta\epsilon$, i.e. $\Gamma_\alpha(\epsilon_\alpha) = \Gamma_\alpha(\beta_\alpha \epsilon_\alpha)$ (e.g. the models~(\ref{eq:exp_rates})
satisfy this hypothesis for $n=0$, while they do not for $n>0$), we find 
that  the COP at maximum power reduces to the universal family of curves
 \begin{equation}
		\mathrm{C}_{\mathrm{op}}(P^{\max}_{[\mathrm{R}]})= 
	\mathrm{C}_{\mathrm{op}}^{(0)} \mathrm{C}_{\mathrm{op}}^{(\mathrm{c})}/
	(1+ \mathrm{C}_{\mathrm{op}}^{(0)} + \mathrm{C}_{\mathrm{op}}^{(\mathrm{c})})\;, 
		\label{eq:cop_maxpow}
\end{equation}
where $\mathrm{C}_{\mathrm{op}}^{(0)}$ represents the COP  when $\beta_\mathrm{H}=\beta_\mathrm{C}$.
It thus follows that for these models the knowledge of 
$\mathrm{C}_{\mathrm{op}}(P^{\max}_{[\mathrm{R}]})$  at a single bath temperature configuration identifies unambiguously the COP for all other temperature differences.
This feature is in contrast with the heat engine mode since, under the same hypothesis, the EMP at arbitrary temperatures depends on the details of the system.

Consider next the maximum power for the models described Eq.~(\ref{eq:exp_rates}). We find that the maximization in Eq.~(\ref{eq:p_generic}) yields $\epsilon^*_\mathrm{H} \to +\infty$ (and a finite value of $\epsilon^*_\mathrm{C}$), which implies
\begin{equation}
    P^{(\max)}_{[\mathrm{R}]} = c_{n} {k_\mathrm{C}}/{\beta_\mathrm{C}^{n+1}}\;,  \qquad 
    \mathrm{C}_{\mathrm{op}}(P^{\max}_{[\mathrm{R}]}) = 0\;, 
    \label{eq:cop_max}
\end{equation}
 where $c_n$ is a dimensionless number which only depends on $n$ for $n>0$, while it is a function of $k_\mathrm{H}/k_\mathrm{C}$ if $n=0$ (see App. D for details). The fact that the corresponding 
 COP is equal to zero is a direct consequence of the divergent value of 
 $\epsilon^*_\mathrm{H}$: physically it means that the maximum cooling power [which is finite, see Eq.~(\ref{eq:cop_max})] is obtained by performing an infinite work, thus by releasing an infinite amount of heat into the hot bath.
 In the more realistic scenario where there are limitations on our control of the
 gaps, say
  $|\epsilon_\alpha|\leq\Delta$,
  the resulting value of $P^{(\max)}_{[\mathrm{R}]}$
will be smaller than in Eq.~(\ref{eq:cop_max}) but the associated COP will be non-zero
with a scaling that for large enough $\Delta$ goes as $\mathrm{C}_{\mathrm{op}}(P^{\max}_{[\mathrm{R}]}) \propto 1/(\beta_\mathrm{C}\Delta)$~(see App. D for details). 
Equation~(\ref{eq:cop_max}) shows that in all models the maximum cooling power only depends on the temperature $1/\beta_\mathrm{C}$ of the cold lead  as a simple power law, and it vanishes as $1/\beta_\mathrm{C} \to 0$. Intuitively this makes sense since it is harder to refrigerate a colder bath and at $1/\beta_\mathrm{C} = 0$ there is no energy to extract from the bath. Furthermore, for $n>0$ the properties of the hot bath (i.e. temperature and coupling constant) do not enter the  $P^{(\max)}_{[\mathrm{R}]}$ formula.   

\begin{figure}[!tb]
	\centering
	\includegraphics[width=0.5\columnwidth]{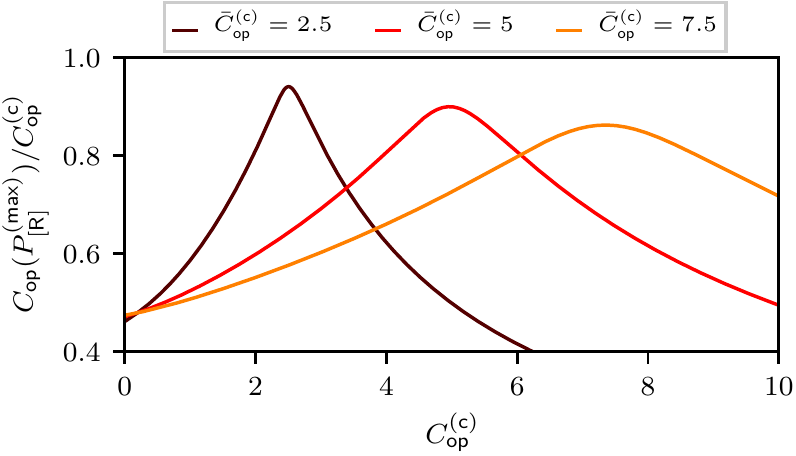}
	\caption{
	 $C_{\mathrm{op}}(P^{(\max)}_{[\mathrm{R}]})$ as a function of $C_{\mathrm{op}}^{(c)}$ (varied by fixing $\beta_\mathrm{H}$ and sweeping over $\beta_\mathrm{C}$), computed using the same Lorentzian filtering rates discussed in Fig.~\ref{fig:eta_max_pow}.
	 Fixing $\sigma=0.01$ and $\bar{\epsilon}_\mathrm{C}=1$ as in Fig.~\ref{fig:eta_max_pow}c, we choose $\bar{\epsilon}_\mathrm{H}/\bar{\epsilon}_\mathrm{C} =$ $7/5$, $6/5$ and $17/15$, corresponding to bath temperature configurations $\bar{C}_{\mathrm{op}}^{(\mathrm{c})} =$ $2.5$, $5$ and $7.5$.  
	  Energies are expressed in units of $1/\beta_\mathrm{H}$ and the CMP does not depend on $\gamma$.
	 }
	\label{fig:cop_max_pow}
\end{figure}

We now return to the possibility of having the CMP arbitrarily close to $\mathrm{C}_{\mathrm{op}}^{(c)}$. As in the heat engine case, from the second equality of Eq.~(\ref{QUESTAQUI}) we see that, if the maximization in Eq.~(\ref{eq:p_generic}) yields values of $\epsilon^*_\mathrm{H}$ and $\epsilon^*_\mathrm{C}$ such that $\epsilon_\mathrm{H}^*\beta_\mathrm{H}\approx \epsilon_\mathrm{C}^*\beta_\mathrm{C}$, then $\mathrm{C}_{\mathrm{op}}(P^{\max}_{[\mathrm{R}]}) \approx \mathrm{C}_{\mathrm{op}}^{(c)}$. Indeed, as we can see in Fig.~\ref{fig:cop_max_pow}, we are able to have a CMP close to $\mathrm{C}_{\mathrm{op}}^{(\mathrm{c})}$ at any desired temperature configuration $\bar{C}_{\mathrm{op}}^{(\mathrm{c})}$ by considering appropriately tuned Lorentzian rates (described in Fig.~\ref{fig:eta_max_pow}).

\section{Thermal accelerator $[\rm A]$ and heater $[\rm H]$ modes.} 
For the  physical models described in Eq.~(\ref{eq:exp_rates}) it turns out that in order to maximize the heat entering the cold bath, it is more convenient to release heat into both baths ($J_\mathrm{H},J_\mathrm{C} < 0$), rather than extracting heat from the hot bath H and releasing it into the cold bath ($J_\mathrm{H}>0,J_\mathrm{C} < 0$). The thermal accelerator mode $[\rm A]$ thus appears  to be useless if we are just interested in maximizing the heat delivered to the cold bath. Accordingly, in the following we shall 
 focus on the heater  $[\rm H]$ mode only with a single bath (or equivalently with two baths at the same temperature).  Assuming 
 to have some physical limit $|\epsilon|\leq \Delta$ on the way we can control 
 the gap, from  Eq.~(\ref{eq:p_generic}) we find  
\begin{equation}
	P_{[\mathrm{H}]}^{(\max)} = \frac{k \Delta^{n+1}}{2}\times \left\{  \begin{array}{lcr}  \tanh{\frac{\beta\Delta}{2}}, &&\mbox{($\mathrm{F}_n$ model)}, \\  1, &&\mbox{($\mathrm{B}_n$ model)},\end{array} \right.
	\label{PA} 
\end{equation}
where $k$ is the coupling constant 
appearing in Eq.~(\ref{eq:exp_rates}). 
Equation~(\ref{PA}) shows that 
the maximum power diverges as $\Delta \to +\infty$, the exponent of $\Delta$ depending on the density of states associated with the rates. Interestingly, the maximum power that can be delivered to the bath vanishes for high temperatures ($\beta\Delta \ll 1$) in the fermionic models, while it is finite and insensitive to temperature in the bosonic models. This is due to the peculiar rates of the bosonic models which diverge for $\beta\epsilon \ll 1$. On the contrary, for low temperatures ($\beta\Delta \gg 1$) both  models yield the same value of $P_{[\mathrm{H}]}^{(\max)}$.

\section{Finite-Time Corrections.} 
The derivation of our main equation 
~(\ref{eq:p_generic}) was obtained under the implicit assumption that one could implement 
infinitesimal control cycles.
Yet this hypothesis  is not as crucial as it may appear. Indeed 
the feasibility of an infinitesimal Otto cycle relies on the ability of performing a very fast driving with respect to the typical time scales of the dynamics, a regime that can be achieved in several experimental setups \cite{Koski2014,Maillet2018}. Furthermore 
 by taking the  square-wave protocol shown in 
Fig.~\ref{fig:main}b characterized by finite time
 intervals  $\tau_\mathrm{H}$ and $\tau_\mathrm{C}$  still fulfilling the ratio $\tau_\mathrm{H}/\tau_\mathrm{C}= \sqrt{{\Gamma_\mathrm{C}(\epsilon_{\rm C})}/{\Gamma_\mathrm{H}(\epsilon_{\rm H})}}$, we find that, at leading order in $dt$,
the maximum power $P^{(\max)}_{[\nu]}(dt)$
only different from the ideal value $P^{(\max)}_{[\nu]}$ of Eq.~(\ref{eq:p_generic}) by a quadratic correction, i.e. 
 $P^{(\max)}_{[\nu]}(dt) \approx (1-  \widetilde{\Gamma}_\mathrm{H} \widetilde{\Gamma}_\mathrm{C}dt^2/12 )P^{(\max)}_{[\nu]}$, 
 where $\widetilde{\Gamma}_\alpha = (\widetilde{\Gamma}\Gamma_\alpha)^{1/2}$, $\widetilde{\Gamma}=\Gamma_\mathrm{H}\Gamma_\mathrm{C}/(\sqrt{\Gamma_\mathrm{H}}+\sqrt{\Gamma_\mathrm{C}})^2$, and all rates are computed for $\epsilon_\mathrm{H}$ and $\epsilon_\mathrm{C}$ that maximize Eq.~(\ref{eq:p_generic}). 
Besides, even in the regime where $\widetilde{\Gamma}_\mathrm{H}dt, \widetilde{\Gamma}_\mathrm{C}dt \gg 1$, $P^{(\max)}_{[\nu]}(dt)$ can be shown (see App. B for details) to only decrease as $(\widetilde{\Gamma}_\mathrm{H}dt/2)^{-1} + (\widetilde{\Gamma}_\mathrm{C}dt/2)^{-1}$,  implying  that a considerable fraction of $P^{(\max)}_{[\nu]}$ can still be achieved 
 also in this case (e.g. see  Fig.~\ref{fig:main}c where we report the $dt$ dependence of $P^{(\max)}_{[\mathrm{H}]}(dt)$ in the heater mode). On the contrary deviations from Eq. (\ref{eq:p_generic}) due to finite time corrections in the quenches turns out to be more relevant.
These last are first order in the ration between the duration of the quench (now different from $0$) and the period of the protocol $dt\,$ (see App. B for details).

\section{Conclusions.} 
We proved that a cycle switching between two extremal values in the fast-driving regime achieves universally the maximum power and 
the maximum cooling rate (respectively for a working medium 
operating as a heat engine or as a refrigerator), regardless of the specific dissipation rates, and we found a general expression for
the external control during the cycle.
The power advantage of modulating the control fields with
rapid adiabatic transformations has 
been observed in the literature
\cite{Feldmann2000,Karimi2016,VanHorne2018} for some specific model and this intuition is in agreement 
with our general results. We also found that the first coefficient of the expansion in power of $\eta_C$ of the EMP
is universal while the second one is linked to the symmetry of the dissipation coefficients.
This paper enlights that the features mentioned above are valid also 
strongly out of equilibrium, while already proven in
low dissipation \cite{Esposito2010} and steady state \cite{Benenti2017} heat engines.
If the bath spectral densities can be suitably tailored through energy filters (as for instance
in \cite{Wegewijs1999}) our protocol allows to reach the
Carnot bound at maximum power both operating as a heat engine or refrigerator, although at the cost of a vanishing power. 
This observation proves the lack of universal upper bounds to the efficiency at maximum power.
Finally, a new scaling for the COP of a bath with flat spectral density is shown and 
a clear dependence of the EMP and the COP at maximum power on the spectral densities of the two 
thermal baths is established.
The results are discussed in detail for some specific models,
from flat bosonic and fermionic baths to environments with
more complicated spectral densities, and finite driving speed effects are analyzed.

\section{Acknowledgments.} 
We thank G. M. Andolina for useful discussions. This work has been supported by SNS-WIS joint
lab ``QUANTRA'', by the SNS internal projects ``Thermoelectricity in nano-devices'', and by the CNR-CONICET
cooperation programme ``Energy conversion in quantum, nanoscale, hybrid devices''.

\appendix

\section*{Appendix A. Optimality of infinitesimal Otto cycles}
 \setcounter{section}{1}
\setcounter{equation}{0}
 
 \label{app_a}
In this appendix we present explicit proof that infinitesimal Otto cycles
are optimal for reaching maximum power performances for our two-level setting.

As a preliminary result we clarify that under periodic modulations of the control
parameters, the master equation (Eq.~(4) of the main text) produces solutions
which asymptotically are also periodic. 
For this purpose let us write Eq.~(4) of the main text
as $\dot{p}(t) = A(t) p(t) +B(t)$ where, for ease of notation, we introduced the functions
\begin{eqnarray} \label{FFD1}
\fl A(t) =- \sum_{\alpha={\rm H,C}}  \lambda_\alpha(t)\Gamma_\alpha[\epsilon(t)]\;,  \qquad 
B(t) =  \sum_{\alpha={\rm H,C}}  \lambda_\alpha(t)\Gamma_\alpha[\epsilon(t)]
p_{\mathrm{eq}}^{(\alpha)}[\epsilon(t)]\;,
\end{eqnarray} 
 and consider periodical driving forces such that
$A(t+ \tau) = A(t)$, $B(t+ \tau)= B(t)$ for all $t$.
By explicitly integration we get
\begin{equation}  p(t) = \int_0^t e^{\int_{t'}^t A(t^{\prime\prime}) dt^{\prime\prime}} B(t') dt' + e^{\int_{0}^t A(t') dt'} p(0). \label{solform}  \end{equation}
Decompose then the integral on the right hand side as
\begin{equation}  
\fl \int\limits_0^t e^{\int_{t'}^t A(t^{\prime\prime}) dt^{\prime\prime}} B(t')dt^\prime =  \int\limits_{t - \tau}^{t}  e^{\int_{t'}^t A(t^{\prime\prime}) dt^{\prime\prime}} B(t')dt^\prime
+ \int\limits_0^{t- \tau}  e^{\int_{t'}^{t- \tau}  A(t^{\prime\prime}) dt^{\prime\prime}} e^{\int_{t-\tau}^{t}  A(t^{\prime\prime}) dt^{\prime\prime}} B(t') dt'.   
\label{period}  
\end{equation}
Notice now that, since $A(t)$ and $B(t)$ are periodic, the quantity $ c(t) = \int_{t - \tau}^{t}  e^{\int_{t'}^t A(t^{\prime\prime}) dt^{\prime\prime}} B(t')dt^\prime$ is also
periodic with period $\tau$, while $d =  e^{\int_{t-\tau}^{t}  A(t^{\prime\prime}) dt^{\prime\prime}}$ is constant in time.
Substituting the previous definitions in Eq. (\ref{period}) and then in Eq. (\ref{solform}) we find
\begin{equation}   p(t) = c(t) + d \int_0^{t- \tau}  e^{\int_{t'}^{t- \tau}  A(t^{\prime\prime}) dt^{\prime\prime}} B(t') dt'  + e^{\int_{0}^t A(t') dt'} p(0). \label{solform2} \end{equation}
In the asymptotic limit  where  the initial condition $p(0)$ has been completely 
forgotten (since  $A(t)  \leq 0$ at all times, the contribution of the initial condition decays exponentially), Eq.~(\ref{solform}) gives
\begin{equation}  p(t-\tau) \approx \int_0^{t- \tau}  e^{\int_{t'}^{t- \tau}  A(t^{\prime\prime}) dt^{\prime\prime}} B(t') dt', \end{equation}
which  substituted in Eq. (\ref{solform2}) allows us to write 
\begin{equation} p(t) \approx  c(t) + d p(t- \tau), \label{periodp} \end{equation}
where we neglected again the contribution coming from the initial condition.
Equation (\ref{periodp}) defines a recursive succession, with limit point equal to $c(t)/(1-d)$, the periodicity of 
$c(t)$ concludes the proof.
This result can also be framed in the general context of Floquet theory~\cite{Chicone1999}.
The Floquet theorem states that a fundamental matrix solution 
of a first order differential equation with periodically driven coefficients is
quasi-periodical, namely can be written as $y(t)= P(t) e^{M t}$ where $P(t)$ is a periodic matrix function (with the same period of the 
coefficients) and $e^{M t}$ is the so called
{\it monodromy matrix}.
The real parts of the eigenvalues of $M$ are responsible of the asymptotic behavior of the solutions and are known as Lyapunov exponents, a stable 
cyclic solution is characterized by their negativity.
In the case of Eq.~(4) of the main text, our calculations reveal that the monodromy matrix is given by the constant
$d$, the sign of the Lyapunov exponent is given by $ \log d < 0$, confirming our predictions about the stability.
\newline

\begin{figure}[t!]
\centering
\includegraphics[scale=.99]{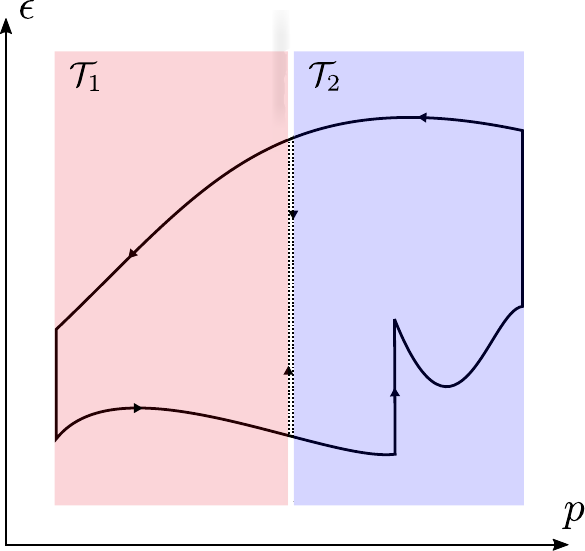}
\caption{The original cycle is represented by a black line following a closed path in the 
$(p, \epsilon)$ plane. The two sub-cycles are the portions of the original one enclosed in the 
light red and in the light blue squares, respectively denoted with $\mathcal{T}_1$ and $\mathcal{T}_2$.}\label{Fig:subcicli}
\end{figure}

In the above paragraph we showed that the asymptotic solution of 
Eq.~(4) of the main text is periodic with the same period of the 
external driving $\epsilon(t)$.
Notice that in the equilibrium scenario the previous result is trivial, since the population instantly relaxes to the Gibbs state that is a monotonic function of the control parameter $\epsilon$.
In our case we can establish only that $p(t)$ and $\epsilon(t)$ share the same period, although finding the proper functional relation between the 
two is absolutely non trivial (cfr. for example \cite{Cavina2017bis}).
However we don't need any additional information to prove that any cycle that is not infinitesimal, namely a square wave protocol in which the 
controls jump at a time much faster that the typical dynamical scale $\Gamma$, cannot achieve the maximum power.
The proof is outlined in the following:
since $\epsilon(t)$ and $p(t)$ share the same periodicity, a cycle can be represented in the $(p,\epsilon)$ plane as a closed curve.
Let us suppose that the optimal cycle $\mathcal{T}$ is not infinitesimal, for example as in Fig.~\ref{Fig:subcicli}. Thus, it is possible to perform an 
instantaneous quench,
for example, in the middle (where the probability is halfway between the minimum and maximum value), and divide the transformation in two 
smaller sub-cycles $\mathcal{T}_1$ and $\mathcal{T}_2$ (cfr. Fig \ref{Fig:subcicli}).
Since the quenches are instantaneous, they don't contribute to the heat exchanged and to the time duration of the process. Furthermore, performing the two 
sub-cycles in series effectively builds a transformation with the same average power of the original cycle, a property that in symbols we can 
exemplify as $P(\mathcal{T}) = P(\mathcal{T}_1 \circ \mathcal{T}_2)$.
Simple calculations reveal that the power of the single sub-cycles cannot be both greater or smaller than the power of the original one, thus
we are left with two possibilities, $P(\mathcal{T}_1) \leq P(\mathcal{T}_1 \circ \mathcal{T}_2) \leq P(\mathcal{T}_2)$ 
or $P(\mathcal{T}_2) \leq P(\mathcal{T}_1 \circ \mathcal{T}_2) \leq P(\mathcal{T}_1)$.
In both cases the original cycle is sub-optimal,  that is absurd, unless $ P(\mathcal{T}_1) = P(\mathcal{T}_1 \circ \mathcal{T}_2) = P(\mathcal{T}_2)$ but even in this case 
we can choose one of the two sub-cycles still preserving optimality.

The previous argument shows that the only candidates for power maximization are those cycles that cannot be divided with a quench as done
in the above proof, thus being infinitesimal. Notice that the previous proof strongly relies on the possibility of performing effectively instantaneous quenches, a characteristic that is better
analyzed in the next appendix. 
At last, by using Pontryagin's minimum principle, it can be shown that if coupling constants $\lambda_\mathrm{H}(t)$ and $\lambda_\mathrm{C}(t)$ fulfill a ``trade-off relation'' (i.e. if one increases, the other one decreases), then the optimal cycle will have $\lambda_\mathrm{C}(t)=0$ and $\lambda_\mathrm{H}(t)=1$, or $\lambda_\mathrm{C}(t)=0$ and $\lambda_\mathrm{H}(t)=1$ at all times \cite{Cavina2017bis}. This implies that the coupling to the baths must be switched during the quenches of the infinitesimal Otto-cycle.

\section*{Appendix B. Maximum Power Formula and Finite-Time Corrections}
 \setcounter{section}{2}
 \setcounter{equation}{0}
 \label{app_b}
In this appendix we prove Eq.~(5) of the main text and discuss the finite-time corrections.

As as shown in the previous appendix, the optimal cycle \textit{must} be an infinitesimal Otto cycle, so we consider a protocol (depicted in Fig.~1b of the main text) where $\epsilon(t)=\epsilon_\mathrm{H}$, $\lambda_\mathrm{H}=1$ and $\lambda_\mathrm{C}=0$ for $t \in [0,\tau_\mathrm{H}]$, and $\epsilon(t)=\epsilon_\mathrm{C}$, $\lambda_\mathrm{H}=0$ and $\lambda_\mathrm{C}=1$ for $t \in [\tau_\mathrm{H}, \tau_\mathrm{H}+\tau_\mathrm{C}]$. The optimal cycle and corresponding power will then be found by taking the limit $dt = \tau_\mathrm{H}+\tau_\mathrm{C}\to 0$ and maximizing over the free parameters $\epsilon_\mathrm{H}$, $\epsilon_\mathrm{C}$ and $\tau_\mathrm{H}/\tau_\mathrm{C}$.

We proceed the following way: first we perform an exact calculation, for arbitrary $\tau_\mathrm{H}$ and $\tau_\mathrm{C}$, of the heat rates $\braket{J_\mathrm{H}}$, $\braket{J_\mathrm{C}}$, averaged over one period, flowing out of the hot and cold  bath respectively. Then, in the limit $dt\to 0$, we find the ratio $\tau_\mathrm{H}/\tau_\mathrm{C}$ that maximizes the power and we find the corresponding expression of the maximum power, proving Eq.~(5) of the main text and the optimal ratio
condition
\begin{eqnarray}\label{CAVINAGOLDENRULE} 
{\tau_\mathrm{H}}/{\tau_\mathrm{C}} = \sqrt{{\Gamma_\mathrm{C}(\epsilon_{\rm C})}/{\Gamma_\mathrm{H}(\epsilon_{\rm H})}}\;.\end{eqnarray} 

The instantaneous heat currents can be written in terms of the probability $p(t)$ by plugging the solution of Eq.~(4) of the main text into Eq.~(3) of the main text.
We denoted with $p_\mathrm{H}(t)$ and $p_\mathrm{C}(t)$ the solution of Eq.~(4) of the main text respectively in the time intervals $\mathcal{I}_\mathrm{H} = [0,\tau_\mathrm{H}]$ and $\mathcal{I}_\mathrm{C} = [\tau_\mathrm{H},\tau_\mathrm{H}+\tau_\mathrm{C}]$. Since the control parameters (i.e. $\epsilon(t)$, $\lambda_\mathrm{H}(t)$ and $\lambda_\mathrm{C}(t)$) are constant in each interval, we have that
\begin{eqnarray}
    &p_\mathrm{H}(t) = He^{-\Gamma_\mathrm{H}t} + p_{\mathrm{eq}}^{(\mathrm{H})},
    &p_\mathrm{C}(t) = Ce^{-\Gamma_\mathrm{C}t} + p_{\mathrm{eq}}^{(\mathrm{C})},
    \label{eq:ph_pc}
\end{eqnarray}
where $H$ and $C$ are two constants and where, for ease of notation, we introduced the
symbols $\Gamma_\alpha := \Gamma_\alpha(\epsilon_\alpha)$ and $p_{\mathrm{eq}}^{(\alpha)} := p_{\mathrm{eq}}^{(\alpha)}(\epsilon_\alpha)$ (for $\alpha=\mathrm{H}, \mathrm{C}$). We determine the two constants $H$ and $C$ by imposing that the probability $p(t)$ is continuous in $t=\tau_\mathrm{H}$, i.e.
\begin{equation}
    p_\mathrm{H}(\tau_\mathrm{H}) = p_\mathrm{C}(\tau_\mathrm{H})
    \label{eq:c1}
\end{equation}
and that $p(t)$ is periodic with period $\tau_\mathrm{H} + \tau_\mathrm{C}$, i.e.
\begin{equation}
    p_\mathrm{H}(0) = p_\mathrm{C}(\tau_\mathrm{H}+\tau_\mathrm{C}).
    \label{eq:c2}
\end{equation}
We impose periodic boundary conditions because, as discussed in the previous appendix, a periodic protocol produces a periodic $p(t)$ after an initial transient time, and we are indeed interested in the ``asymptotic'' regime. Equations~(\ref{eq:c1}) and (\ref{eq:c2}) reduce to the following linear-algebra problem for the constants $H$ and $C$:
\begin{equation}
    \frac{1}{p_{\mathrm{eq}}^{(\mathrm{C})} - p_{\mathrm{eq}}^{(\mathrm{H})}}
    \left(\matrix{
        e^{-\Gamma_\mathrm{H}\tau_\mathrm{H}} & -e^{-\Gamma_\mathrm{C}\tau_\mathrm{H}} \cr
        1 & -e^{-\Gamma_\mathrm{C}(\tau_\mathrm{H}+\tau_\mathrm{C})} 
        }\right)
    \left(\matrix{
        H \cr C
     }\right)
    = 
    \left(\matrix{
        1 \cr 1
    }\right),
    \label{eq:hc_mat} 
\end{equation}
with solution 
\begin{equation}
    \left(\matrix{
        H \cr C
    }\right)
    =\frac{\left( p_{\mathrm{eq}}^{(\mathrm{C})} - p_{\mathrm{eq}}^{(\mathrm{H})} \right)}{ \sinh{\left[ (\Gamma_\mathrm{H}\tau_\mathrm{H} + \Gamma_\mathrm{C}\tau_\mathrm{C})/2 \right]} }
    \left(\matrix{
        e^{\Gamma_\mathrm{H}\tau_\mathrm{H}/2} \sinh{\left(\Gamma_\mathrm{C}\tau_\mathrm{C}/2\right)} \cr
        -e^{\Gamma_\mathrm{C}\tau_\mathrm{H}} e^{\Gamma_\mathrm{C}\tau_\mathrm{C}/2} \sinh{\left( \Gamma_\mathrm{H}\tau_\mathrm{H}/2 \right)}
    }\right),
    \label{eq:hc_explicit}
\end{equation}
which, via Eq.~(\ref{eq:ph_pc}), completely determine $p(t)$. 
By substituting Eq.~(\ref{eq:ph_pc}) into  Eq.~(3) of the main text, we can write the averaged heat rates $\braket{J_\mathrm{H}}$ and $\braket{J_\mathrm{C}}$ as
\begin{eqnarray}
    \fl &\braket{J_\mathrm{H}} :=\frac{1}{\tau_\mathrm{H}+\tau_\mathrm{C}} \int\limits_{\mathcal{I}_\mathrm{H}} J_\mathrm{H}dt = \frac{\epsilon_\mathrm{H}}{\tau_\mathrm{H}+\tau_\mathrm{C}} \int\limits_{\mathcal{I}_\mathrm{H}} \dot{p}_\mathrm{H}dt  = \frac{\epsilon_\mathrm{H}H}{\tau_\mathrm{H}+\tau_\mathrm{C}} \left[ e^{-\Gamma_\mathrm{H}\tau_\mathrm{H}} -1 \right], \nonumber \\
    \fl &\braket{J_\mathrm{C}} :=\frac{1}{\tau_\mathrm{H}+\tau_\mathrm{C}} \int\limits_{\mathcal{I}_\mathrm{C}} J_\mathrm{C}dt = \frac{\epsilon_\mathrm{C}}{\tau_\mathrm{H}+\tau_\mathrm{C}} \int\limits_{\mathcal{I}_\mathrm{C}} \dot{p}_\mathrm{C}dt =  \frac{\epsilon_\mathrm{C}C}{\tau_\mathrm{H}+\tau_\mathrm{C}} e^{-\Gamma_\mathrm{C}\tau_\mathrm{H}}\left[ e^{-\Gamma_\mathrm{C}\tau_\mathrm{C}} -1 \right],
\label{eq:ph_pc_deriv}
\end{eqnarray}
where we use the fact that $\epsilon(t)$ is constant in each $\mathcal{I}_\alpha$ and  the fact that, since the two-level system is coupled to one bath at a time, $\dot{p}_\alpha=\dot{p}$ with $p(t)=p_\mathrm{H}(t)$ during $\mathcal{I}_\mathrm{H}$ and $p(t)=p_\mathrm{C}(t)$ during $\mathcal{I}_\mathrm{C}$. Using the expressions for $H$ and $C$ given in Eq.~(\ref{eq:hc_explicit}), we can rewrite Eq.~(\ref{eq:ph_pc_deriv}) as 
\begin{equation}
    \fl\braket{J_\mathrm{H/C}} = \pm \frac{\epsilon_{\mathrm{H/C}}}{\tau_\mathrm{H}+\tau_\mathrm{C}} \frac{\Gamma_\mathrm{H}\tau_\mathrm{H}\Gamma_\mathrm{C}\tau_\mathrm{C}}{\Gamma_\mathrm{H}\tau_\mathrm{H}+\Gamma_\mathrm{C}\tau_\mathrm{C}} \left( p_{\mathrm{eq}}^{(\mathrm{H})} - p_{\mathrm{eq}}^{(\mathrm{C})} \right)
    \frac{ \left( \Gamma_\mathrm{H}\tau_\mathrm{H}/2\right)^{-1} + \left( \Gamma_\mathrm{C}\tau_\mathrm{C}/2 \right)^{-1} }{ \coth{\left( \Gamma_\mathrm{H}\tau_\mathrm{H}/2 \right)} + \coth{\left( \Gamma_\mathrm{C}\tau_\mathrm{C}/2 \right)} }.
    \label{eq:p_dt_nomax}
\end{equation}
We now impose that $dt=\tau_\mathrm{H}+\tau_\mathrm{C}$ by setting $\tau_\mathrm{H} = \theta dt$ and $\tau_\mathrm{C}=(1-\theta)dt$, for $\theta \in [0,1]$, in Eq.~(\ref{eq:p_dt_nomax}). Taking 
hence the infinitesimal cycle limit 
$dt\to 0$ we get 
\begin{equation}
    \braket{J_\mathrm{H/C}} = \pm \epsilon_{\mathrm{H/C}} \frac{\Gamma_\mathrm{H}\theta \; \Gamma_\mathrm{C}(1-\theta)}{\Gamma_\mathrm{H}\theta+\Gamma_\mathrm{C}(1-\theta)} \left( p_{\mathrm{eq}}^{(\mathrm{H})} - p_{\mathrm{eq}}^{(\mathrm{C})} \right).
    \label{eq:p_dt_theta}
\end{equation}
The maximization over $\theta$ of the above expression yields the condition 
\begin{equation}
    \frac{\theta}{1-\theta} = \sqrt{\frac{\Gamma_\mathrm{C}}{\Gamma_\mathrm{H}}},
    \label{eq:theta_max}
\end{equation}
which, multiplying by $dt$ the numerator and the denominator of the left hand side of Eq.~(\ref{eq:theta_max}), proves Eq.~(\ref{CAVINAGOLDENRULE}). 
Solving hence  Eq.~(\ref{eq:theta_max}) for $\theta$ and plugging the result into Eq.~(\ref{eq:p_dt_theta}) yields
\begin{equation}
    \braket{J_\mathrm{H/C}} = \pm \epsilon_{\mathrm{H/C}} \frac{\Gamma_\mathrm{H}\Gamma_\mathrm{C}}{ \left(\sqrt{\Gamma_\mathrm{H}} +\sqrt{\Gamma_\mathrm{C}} \right) ^2} \,  \left( p_{\mathrm{eq}}^{(H)} - p_{\mathrm{eq}}^{(C)} \right),
    \label{eq:heat_pow}
\end{equation}
which replaced into Eq.~(1) of the main text, and maximizing with respect to 
the only two free parameters left, i.e. $\epsilon_\mathrm{H}$ and $\epsilon_\mathrm{C}$,
 allows us to derive Eq.~(5) of the main text
for all four thermal machine modes.  
An additional comment has to be made for the accelerator mode [A], that aims at maximizing the heat released into the cold bath while extracting heat from the hot bath. 
By definition, we must restrict the maximization in Eq.~(5) of the main text to guarantee $\langle J_\mathrm{H}\rangle \geq 0$, e.g. by
forcing $\mathcal{C}$ to be 
	$(\epsilon_\mathrm{H}\geq 0 \cap \beta_\mathrm{C}\epsilon_\mathrm{C} \geq \beta_\mathrm{H}\epsilon_\mathrm{H}) \cup (\epsilon_\mathrm{H}\leq 0 \cap \beta_\mathrm{C}\epsilon_\mathrm{C} \leq \beta_\mathrm{H}\epsilon_\mathrm{H})$.
On the other hand, the heater mode consists of heating a {single reservoir} whose interaction with the two-level system is described by a rate $\Gamma(\epsilon)$ and equilibrium probability $p_{\mathrm{eq}}(\epsilon)$. So in this case the maximization must be performed taking  $\Gamma_\alpha(\epsilon) = \Gamma(\epsilon)$ and $p_{\mathrm{eq}}^{(\alpha)}(\epsilon) = p_{\mathrm{eq}}(\epsilon)$ (for $\alpha=$ H, C). If we also require that $\Gamma(\epsilon) = \Gamma(-\epsilon)$, which physically means that the rates do not distinguish which one of the two energy levels is the ground and excited state, we find that Eq.~(5) can be simplified to 
 \begin{equation}
 	P^{(\max)}_{\mathrm{[H]}} = \max_{ \epsilon \geq 0, \, \epsilon \in \mathcal{C}} \frac{1}{2}\,\epsilon\, \Gamma(\epsilon) \left[1 - 2 p_{\mathrm{eq}}(\epsilon)\right] ,
 	\label{eq:p_heater}
 \end{equation}
and the corresponding optimal cycle is given by an Otto cycle where $\tau_\mathrm{H}=\tau_\mathrm{C}$ and the value $\epsilon$ that maximizes Eq.~(\ref{eq:p_heater}) determines $\epsilon_\mathrm{H} = - \epsilon_\mathrm{C} = \epsilon $. Thus the optimal cycle in the heater case corresponds to attempting continuous population inversions.

\subsection{Finite-Time Corrections part one} 
 Setting $\tau_\mathrm{H} = \theta dt$ and $\tau_\mathrm{C}=(1-\theta)dt$ in Eq.~(\ref{eq:p_dt_nomax}), and plugging in the expression of $\theta$ that satisfies Eq.~(\ref{eq:theta_max}), we find that the average heat rate for an arbitrary period $dt$ is given by
\begin{equation}
    \fl\braket{J_\mathrm{H/C}(dt)} = \pm \epsilon_{\mathrm{H/C}} \frac{\Gamma_\mathrm{H}\Gamma_\mathrm{C}}{ \left(\sqrt{\Gamma_\mathrm{H}} +\sqrt{\Gamma_\mathrm{C}} \right) ^2}
     \left( p_{\mathrm{eq}}^{(H)} - p_{\mathrm{eq}}^{(C)} \right)
    \frac{( \widetilde{\Gamma}_\mathrm{H} dt/2 )^{-1} + ( \widetilde{\Gamma}_\mathrm{C} dt/2 )^{-1} }{ \coth{( \widetilde{\Gamma}_\mathrm{H} dt/2 )} + \coth{( \widetilde{\Gamma}_\mathrm{C} dt/2 )} },
\end{equation}
where $\widetilde{\Gamma}_\alpha = (\widetilde{\Gamma}\Gamma_\alpha)^{1/2}$ and $\widetilde{\Gamma}=\Gamma_\mathrm{H}\Gamma_\mathrm{C}/(\sqrt{\Gamma_\mathrm{H}}+\sqrt{\Gamma_\mathrm{C}})^2$. Plugging this results into Eq.~(1) of the main text and maximizing over $\epsilon_\mathrm{H}$ and $\epsilon_\mathrm{C}$ yields the expression 
\begin{equation}
	P^{(\max)}_{[\nu]}(dt) =  \frac{( \widetilde{\Gamma}_\mathrm{H} dt/2 )^{-1} + ( \widetilde{\Gamma}_\mathrm{C} dt/2 )^{-1} }{ \coth{( \widetilde{\Gamma}_\mathrm{H} dt/2 )} + \coth{( \widetilde{\Gamma}_\mathrm{C} dt/2 )} }\; P^{(\max)}_{[\nu]},
	\label{eq:p_dt}
\end{equation}
which provides the finite time version of Eq. (5) of the main text. 
On one hand, as anticipated in the main text, by expanding Eq.~(\ref{eq:p_dt}) for small $dt$, we find the following quadratic correction 
\begin{eqnarray} 
P^{(\max)}_{[\nu]}(dt) \approx ( 1-  \widetilde{\Gamma}_\mathrm{H} \widetilde{\Gamma}_\mathrm{C}dt^2/12 ) P^{(\max)}_{[\nu]}\;.\end{eqnarray} 
On the other hand for 
 $\widetilde{\Gamma}_\mathrm{H}dt, \widetilde{\Gamma}_\mathrm{C}dt \gg 1$, we get 
 \begin{eqnarray} 
P^{(\max)}_{[\nu]}(dt) \approx \frac{ (\widetilde{\Gamma}_\mathrm{H} dt/2 )^{-1} + ( \widetilde{\Gamma}_\mathrm{C} dt/2 )^{-1} }{2} \; P^{(\max)}_{[\nu]}  \;,\end{eqnarray} 
implying that a considerable fraction of $P^{(\max)}_{[\nu]}$ can be achieved even if the driving frequency is slower than the typical rate.
Notice that Eq.~(\ref{eq:p_dt}) is a strictly decreasing function of $dt$; this is consistent with the fact that an infinitesimal cycle is indeed the optimal solution.
 
We conclude by observing that we can simplify Eq.~(\ref{eq:p_dt}) for the heater mode where a single reservoir is coupled to the two-level system. Under the hypothesis leading to Eq.~(\ref{eq:p_heater}), we find that
\begin{equation}
 	P^{(\max)}_{\mathrm{[H]}}(dt) = \frac{\tanh{\left(dt \Gamma /4\right)}}{dt \Gamma/4}P^{(\max)}_{\mathrm{[H]}} ,
 	\label{eq:ph_dt}
\end{equation}
where $\Gamma$ is computed in the value of $\epsilon$ that maximizes Eq.~(\ref{eq:p_heater}). Figure~1c of the main text, which is a plot of Eq.~(\ref{eq:ph_dt}), shows that $P^{(\max)}_{\mathrm{[H]}}(dt) \approx P^{(\max)}_{\mathrm{[H]}}$ up to  $dt\Gamma \approx 2$, while for $dt\Gamma = 10 \gg 1$, $P^{(\max)}_{\mathrm{[H]}}(dt)$ is only decreased of a factor two.

\subsection{Finite-Time Corrections part two: the Quenches} \label{app:fquench}
Finite-time corrections to the power may not only arise from the finite duration of the isothermal transformations (i.e. from a finite value of $\tau_\mathrm{C}$ and $\tau_\mathrm{H}$), but  also from a finite duration $\tau$ of the quenches, i.e. of the transformations during which $\epsilon$ changes between the two extremal values $\epsilon_\mathrm{C}$ and $\epsilon_\mathrm{H}$. We will thus assume that each quench is carried out in a time $\tau$. The aim of this appendix is to show how these effects could be accounted for, and to estimate the leading order corrections to the maximum power delivered by the heat engine due to this effect; analogous considerations hold also for the other machines. We will thus restrict ourself to the regime $\tau \ll dt \ll \gamma^{-1}$, where $dt = \tau_\mathrm{C}+\tau_\mathrm{H}$ and $\gamma$ is the characteristic rate of the system during the protocol. The first inequality states that the duration of the quenches is much smaller than the duration of the isothermal transformations.
The second inequality implies that the finite-time corrections discussed in the previous subsection are neglected, since they have been previously discussed.

Using the results of App. A, we know that $p(t)$ has a limit cycle with the same period of $\epsilon(t)$. If we further assume that the protocol is much faster than $\gamma$, the probability tends to a fixed value $\bar{p}$ given by
\begin{equation}
	\bar{p} = \frac{ \int_0^T ds \Gamma(s) p_{\mathrm{eq}}(s) }{\int_0^T ds \Gamma(s)},
	\label{eq:barp}
\end{equation}
where $T = dt+2\tau$ is the total duration of the protocol, $\Gamma(s) = \lambda_\mathrm{H}(s) \Gamma_\mathrm{H}[\epsilon(s)] + \lambda_\mathrm{C}(s) \Gamma_\mathrm{C}[\epsilon(s)]$ and 
\begin{equation}
	p_{\mathrm{eq}}(s) = \frac{\lambda_\mathrm{H}(s) \Gamma_\mathrm{H}[\epsilon(s)]p_{\mathrm{eq}}^{(\mathrm{H})}[\epsilon(s)] + \lambda_\mathrm{C}(s) \Gamma_\mathrm{C}[\epsilon(s)]p_{\mathrm{eq}}^{(\mathrm{C})}[\epsilon(s)]}{ \lambda_\mathrm{H}(s) \Gamma_\mathrm{H}[\epsilon(s)] + \lambda_\mathrm{C}(s) \Gamma_\mathrm{C}[\epsilon(s)]}.
\end{equation}
By using again the hypothesis that the protocol is much faster than $\gamma$, we can write the power of the heat engine, averaged over one period, as
\begin{equation}
	P_{\mathrm{[E]}}(\lambda) = \frac{1}{T} \int_0^T ds\, \epsilon(s) \Gamma(s) \left[ p_{\mathrm{eq}}(s) - \bar{p} \right].
	\label{eq:p_quench_def}
\end{equation}
As in the ideal protocol (see Fig 1b of the main text), we will assume that during the two isothermal transformations we respectively have $\epsilon(s) = \epsilon_\mathrm{H}$, $\lambda_\mathrm{H}(s) =1$, $\lambda_\mathrm{C}(s)=0$ and $\epsilon(s) = \epsilon_\mathrm{C}$, $\lambda_\mathrm{H}(s) =0$, $\lambda_\mathrm{C}(s)=1$. This means that we are coupled to one bath at a time. Instead, during the quenches we assume that all three control parameters ($\epsilon(s)$, $\lambda_\mathrm{H}(s)$ and $\lambda_\mathrm{C}(s)$) vary linearly in time between the corresponding extremal values. We thus divide the integral in Eq.~(\ref{eq:p_quench_def}) in the four different transformations:
\begin{eqnarray}
	\fl P_{\mathrm{[E]}}(\tau) = \frac{\int_0^{\tau_\mathrm{H}} (\dots) + \int_{\tau_\mathrm{H}}^{\tau_\mathrm{H}+\tau} (\dots) + \int_{\tau_\mathrm{H}+\tau}^{\tau_\mathrm{H}+\tau_\mathrm{C}+\tau} (\dots)  + \int_{\tau_\mathrm{H}+\tau_\mathrm{C}+\tau}^{\tau_\mathrm{H}+\tau_\mathrm{C}+2\tau} (\dots) }{dt+2\tau} \equiv \nonumber  \\
	\frac{  \mathcal{W}_{\mathrm{H}} + \mathcal{W}_{\mathrm{H}\to \mathrm{C}} + \mathcal{W}_{ \mathrm{C}} + \mathcal{W}_{\mathrm{C}\to \mathrm{H}} }{dt+2\tau},
	\label{eq:peq1}
\end{eqnarray}
where $(\dots)$ stands for $ds\,\epsilon(s) \Gamma(s) \left[ p_{\mathrm{eq}}(s) - \bar{p} \right]$. In the regime we consider the power, up to leading order corrections in $\tau/dt$, can be written as
\begin{equation}
	P_{\mathrm{[E]}}(\tau) = \frac{  \mathcal{W}_{\mathrm{H}}  + \mathcal{W}_{ \mathrm{C}}}{dt} \left(1 - \frac{2\tau}{dt} \right) + \frac{\mathcal{W}_{\mathrm{H}\to \mathrm{C}} + \mathcal{W}_{\mathrm{C}\to \mathrm{H}}}{dt},
	\label{eq:p_quench_2}
\end{equation}
where the first addend is obtained by means of a first order expansion in $\tau/dt$ of the denominator in the r.h.s. of Eq. (\ref{eq:peq1}).

We wish to compare Eq.~(\ref{eq:p_quench_2}) to the power $P_{\mathrm{[E]}}^{(\max)}$ achieved in the ideal protocol, so we will estimate the four terms $\mathcal{W}_{\mathrm{H}}$, $\mathcal{W}_{\mathrm{C}}$, $\mathcal{W}_{\mathrm{H}\to \mathrm{C}}$ and $\mathcal{W}_{\mathrm{C}\to \mathrm{H}}$. First, we notice that $\bar{p}$ depends on the whole protocol, see Eq.~(\ref{eq:barp}). We can thus write
\begin{equation}
	\bar{p} = \bar{p}^{(0)} + \delta \bar{p}^{(1)},
	\label{eq:p_01}
\end{equation}
where $\bar{p}^{(0)}$ is the value of $\bar{p}$ in the ideal protocol, and $\delta \bar{p}^{(1)}$ the corrections due to the finite-time quenches. These two terms can be calculated simply by dividing the integrals in the definition of $\bar{p}$ as we did for $P_{\mathrm{[E]}}(\tau)$. It is easy to see that $\delta \bar{p}^{(1)}$ is of the order $\tau/dt$. Since $\mathcal{W}_{\mathrm{H}}$ and $\mathcal{W}_{\mathrm{C}}$ are linear functions of $\bar{p}$, and $\bar{p}=\bar{p}^{(0)} + \delta \bar{p}^{(1)}$, we have that, for $\alpha = \mathrm{H}, \mathrm{C}$,
\begin{equation}
	\mathcal{W}_{\alpha} = \tau_\alpha \epsilon_\alpha \Gamma_\alpha(\epsilon_\alpha) \left[ p_{\mathrm{eq}}^{(\alpha)}(\epsilon_\alpha) - \bar{p} \right] =
	\mathcal{W}_{\alpha}^{(0)} + \mathcal{W}_{\alpha}^{(1)},
	\label{eq:walpha}
\end{equation}
where $\mathcal{W}_{\alpha}^{(0)}$ is the work extracted in the ideal protocol during the isothermal transformation, while $\mathcal{W}_{\alpha}^{(1)}$ represents the corrections due to the variation in population $\bar{p}$ induced by the finite-time quenches. We have that
\begin{equation}
	\mathcal{W}_{\alpha}^{(1)} =  -\tau_\alpha \epsilon_\alpha \Gamma_\alpha(\epsilon_\alpha) \delta\bar{p}^{(1)} \propto \epsilon_\alpha O(\gamma \tau),
	\label{eq:wa1}
\end{equation}
where the last term means that $\mathcal{W}_{\alpha}^{(1)}$ is of the order $\gamma\tau$. 

Next, we need to estimate $\mathcal{W}_{\mathrm{H}\to \mathrm{C}}$ and $\mathcal{W}_{\mathrm{C}\to \mathrm{H}}$. By inspecting the definition, we see that
\begin{equation}
	\mathcal{W}_{\mathrm{H}\to \mathrm{C}} = \mathcal{W}_{\mathrm{C}\to \mathrm{H}} \propto \braket{\epsilon} O(\gamma\tau),
	\label{eq:whc}
\end{equation}
where $\braket{\epsilon}$ is a characteristic value of the energy gap during the quench.

Now we can return to Eq.~(\ref{eq:p_quench_2}). Using Eqs.~(\ref{eq:walpha}), (\ref{eq:wa1}) and (\ref{eq:whc}), and noticing that the order of magnitude of $\gamma \braket{\epsilon}$ is the same as $P_{\mathrm{[E]}}^{(\max)}$, we find that all the corrections previously discussed are of the order $\gamma/dt$. We thus conclude that
\begin{equation}
	P_{\mathrm{[E]}}(\tau) =  P_{\mathrm{[E]}}^{(\max)} \left[1 - O(\tau/dt) \right] ,
\end{equation}
where the corrections must be negative by virtue of the theorem proved in App. A. The impact of finite time quenches is thus first order $\tau/dt$.

\section*{Appendix C. Efficiency at Maximum Power}
 \setcounter{section}{3}
 \setcounter{equation}{0}
 \label{app_c}

For small temperature differences, i.e. for small values of $\eta_\mathrm{c}$, we can consider an expansion of the efficiency at maximum power of the kind
\begin{equation}
	\eta(P_{[\mathrm{E}]}) = a_1\eta_\mathrm{c} + a_2\eta_\mathrm{c}^2 + \dots.
	\label{eq:etapmax_series_etac}
\end{equation}
In this appendix we prove that $a_1 = 1/2$, while for symmetric or constant rates we further have $a_2=1/8$.
The maximum power of a heat engine (without constraints on the control parameters) can be written as [see Eq.~(5) of the main text]
\begin{equation}
	P_{[\mathrm{E}]}^{(\max)} = \max_{(x_\mathrm{H}, x_\mathrm{C})} P_{[\mathrm{E}]}(x_\mathrm{H},x_\mathrm{C}),
	\label{eq:pe_x}
\end{equation}
where 
\begin{equation}
	 P_{[\mathrm{E}]}(x_\mathrm{H},x_\mathrm{C}) := \frac{g(x_\mathrm{H},x_\mathrm{C}; \, \eta_\mathrm{c})}{\beta_\mathrm{H}}\left[ x_\mathrm{H} - x_\mathrm{C}(1-\eta_\mathrm{c}) \right] \left[ f(x_\mathrm{H}) - f(x_\mathrm{C}) \right],
	\label{eq:p_x}
\end{equation}
$x_\alpha = \epsilon_\alpha \beta_\alpha$ (for $\alpha$ = H, C), $f(x):=[1+\exp{(x)}]^{-1}$ and, expressing the $\Gamma_\alpha$ as a function of the gap $\epsilon$ and of the inverse temperature $\beta_\alpha$ of lead $\alpha$,
\begin{equation}
	g(x_\mathrm{H},x_\mathrm{C};\, \eta_\mathrm{c}) :=  \frac{\Gamma_\mathrm{H}(x_\mathrm{H},\beta_\mathrm{H})\,\Gamma_\mathrm{C}(x_\mathrm{C},\beta_\mathrm{H}/(1-\eta_\mathrm{c}))}{ \left(\sqrt{\Gamma_\mathrm{H}(x_\mathrm{H},\beta_\mathrm{H})} +\sqrt{\Gamma_\mathrm{C}(x_\mathrm{C},\beta_\mathrm{H}/(1-\eta_\mathrm{c}))} \right) ^2}.
	\label{eq:g_def}
\end{equation}
In Eq.~(\ref{eq:g_def}) we decide to express $\beta_\mathrm{C}$ as $\beta_\mathrm{H}/(1-\eta_\mathrm{c})$ because we are interested in performing an expansion in $\eta_\mathrm{c}$ around a single inverse temperature $\beta_\mathrm{H}$.
Let $x^*_\mathrm{H}$ and $x^*_\mathrm{C}$ be respectively the values of $x_\mathrm{H}$ and $x_\mathrm{C}$ that maximize $P_{[\mathrm{E}]}(x_\mathrm{H},x_\mathrm{C})$. By inspecting Eqs.~(\ref{eq:p_x}) and (\ref{eq:g_def}), we see that $x^*_\alpha$ is a function of $\eta_\mathrm{c}$ (and of $\beta_\mathrm{H}$ through $g$), so we can express $x^*_\alpha$ as a power series in $\eta_\mathrm{c}$:
\begin{eqnarray}
	x^*_\mathrm{H} = m_0 + m_1 \eta_\mathrm{c} +  m_2 \eta_\mathrm{c}^2 + \dots, \nonumber \\
	x^*_\mathrm{C} = m_0 + n_1 \eta_\mathrm{c} +  n_2 \eta_\mathrm{c}^2 + \dots.
\label{eq:x_bar}
\end{eqnarray}
Both $x^*_\mathrm{H}$ and $x^*_\mathrm{C}$ have the same leading order term. This can be seen considering Eq.~(\ref{eq:p_x}) at $\eta_\mathrm{c}=0$: $g(x_\mathrm{H},x_\mathrm{C};\,0)/\beta_\mathrm{H}\geq 0$, while $\left[ x_\mathrm{H} - x_\mathrm{C} \right] \left[ f(x_\mathrm{H}) - f(x_\mathrm{C}) \right]\leq 0$, so the maximum power is zero (at equal temperatures, the second law forbids the possibility of extracting work). Inspecting Eq.~(\ref{eq:p_x}), it is easy to see that zero power at $\eta_\mathrm{c}=0$ implies $x_\mathrm{H} = x_\mathrm{C}$.
Using Eq.~(7) of the main text, we have that
\begin{equation}
	\eta(P_{[\mathrm{E}]}^{(\max)}) = 1 - \frac{x^*_\mathrm{C}}{x^*_\mathrm{H}}\left(1 - \eta_\mathrm{c} \right),
	\label{eq:eff_x}
\end{equation}
so plugging Eq.~(\ref{eq:x_bar}) into Eq.~(\ref{eq:eff_x}) and expressing $\eta(P_{[\mathrm{E}]}^{(\max)})$ as a power series in $\eta_\mathrm{c}$, we find that
\begin{equation}
	\eta(P_{[\mathrm{E}]}^{(\max)}) = (1+b_1)\eta_\mathrm{c} + \frac{1}{2}(1+b_2)\eta_\mathrm{c}^2,
	\label{eq:eta_max_exp}
\end{equation}
where 
\begin{eqnarray}
	b_1 = \frac{m_1-n_1}{m_0}, 
	\qquad b_2 = \frac{m_1}{m_0} + 2 \frac{m_2-n_2}{m_0}. \nonumber
\end{eqnarray}
Thus, the knowledge of $b_1$ and $b_2$ implies also the knowledge of $a_1$ and $a_2$. Also the maximum power $P_{[\mathrm{E}]}(x^*_\mathrm{H}, x^*_\mathrm{C})$ can be written as a power series in $\eta_\mathrm{c}$ by plugging the expansion Eq.~(\ref{eq:x_bar}) into Eq.~(\ref{eq:p_x}). This yields
\begin{equation}
	P_{[\mathrm{E}]}(x^*_\mathrm{H}, x^*_\mathrm{C}) = \frac{1}{\beta_\mathrm{H}} \sum_{n=0}^{+\infty} P_{[\mathrm{E}]}^{(n)}\eta_\mathrm{c}^n,
\end{equation}
where the coefficients $ P_{[\mathrm{E}]}^{(n)}$ are functions of $m_i$, $n_i$ (for $i=0,1,2,\dots$) and of $\beta_\mathrm{H}$. We now wish to determine $b_1$ and $b_2$ by maximizing $P_{[\mathrm{E}]}^{(n)}$, starting from the lowest orders. We find that $P_{[\mathrm{E}]}^{(0)} = P_{[\mathrm{E}]}^{(1)} = 0$ and 
\begin{equation}
	P_{[\mathrm{E}]}^{(2)} =  \left[ -\frac{b_1(1+b_1)}{2} \right] \frac{m_0^2 \, g(m_0,m_0;\, 0)}{1+\cosh{m_0}},
	\label{eq:p2}
\end{equation}
where we expressed $n_1$ in terms of $b_1$. The last fraction in Eq.~(\ref{eq:p2}) is positive, so $P_{[\mathrm{E}]}^{(2)}$ is maximized by choosing $b_1$ that maximizes the term in square brackets, and $m_0$ that maximizes the last fraction. The maximization of the first term yields $b_1=-1/2$, which readily implies [see Eq.~(\ref{eq:eta_max_exp})] $a_1 = 1/2$, as we wanted to prove. The maximization of the second term allows us to find the following implicit expression for $m_0$ 
\begin{equation}
	\fl g(m_0,m_0;\,0)\left[ 2 - m_0\tanh{\left( \frac{m_0}{2} \right)} \right] + m_0\left[ \partial_{x_\mathrm{H}} g(m_0,m_0;\,0) + \partial_{x_\mathrm{C}} g(m_0,m_0;\,0) \right] = 0,
	\label{eq:a0}
\end{equation}
where $ \partial_{x_\alpha} g(m_0,m_0;\,0)$ denotes the partial derivative of $g(x_\mathrm{H},x_\mathrm{C};\,\eta_\mathrm{c})$, respect to $x_\alpha$, calculated in $x_\mathrm{H} = x_\mathrm{C} = m_0$ and $\eta_\mathrm{c}=0$.
In order to compute $b_2$, we must maximize also higher order terms of the power. It turns out that $P_{[\mathrm{E}]}^{(3)}$ only depends on $m_0$ if we impose that $b_1=-1/2$ and that $m_0$ satisfies Eq.~(\ref{eq:a0}). Thus, there is nothing to optimize, so we must analyze the next order. $P_{[\mathrm{E}]}^{(4)}$ is a function of $m_0$, $m_1$, $n_1$, $m_2$, $n_2$ and $\beta_\mathrm{C}$.
We write $m_1$ in terms of $b_2$, which is the only coefficient that determines $a_2$. We further express $n_1$ in terms of $b_1$, and impose $b_1=-1/2$. At last, we write $g(m_0,m_0;\,0)$ in terms of its partial derivatives using Eq.~(\ref{eq:a0}). This leads to an expression of $P_{[\mathrm{E}]}^{(4)}$ as a function of $m_0$ (which is implicitly known), $b_2$, $m_2$, $n_2$ and $\beta_\mathrm{H}$.
We maximize $P_{[\mathrm{E}]}^{(4)}$ by setting to zero both partial derivatives of $P_{[\mathrm{E}]}^{(4)}$ respect to $b_2$ and $m_2$. We thus find the following expression for $b_2$:
\begin{equation}
	b_2 = \frac{ m_0 \tanh{\left(\frac{m_0}{2}\right)}}{8} \cdot \frac{ \partial_{x_\mathrm{H}} g - \partial_{x_\mathrm{C}} g  }{  \partial_{x_\mathrm{H}} g +  \partial_{x_\mathrm{C}} g} - \frac{ 2 \partial_{x_\mathrm{H}} g +  \partial_{x_\mathrm{C}} g }{2\left(  \partial_{x_\mathrm{H}} g +  \partial_{x_\mathrm{C}} g \right)},
	\label{eq:c3}
\end{equation}
where all partial derivatives of $g$ are computed in $x_\mathrm{H}=x_\mathrm{C}=m_0$ and $\eta_\mathrm{c}=0$. This is, in principle, a closed expression for $b_2$, thus for $a_2$, since $m_0$ is defined in Eq.~(\ref{eq:a0}), and Eq.~(\ref{eq:c3}) only depends on $m_0$. Eq.~(\ref{eq:c3}) shows that in general $b_2$, thus $a_2$, will depend on the specific rates. However, if $ \partial_{x_\mathrm{H}} g =  \partial_{x_\mathrm{C}} g$, the first term in Eq.~(\ref{eq:c3}) vanishes, while the second one reduces to a number, yielding $b_2=-3/4$. Indeed, plugging this value of $b_2$ into Eq.~(\ref{eq:eta_max_exp}) yields precisely $a_2=1/8$. We conclude the proof by noticing that if the rates are symmetric, i.e. $\Gamma_\mathrm{H}(\epsilon,\beta) = \Gamma_\mathrm{C}(\epsilon,\beta)$, $g(x_\mathrm{H}, x_\mathrm{C};\,0)$ is a symmetric function upon exchange of $x_\mathrm{H}$ and $x_\mathrm{C}$. This implies that $\partial_{x_\mathrm{H}} g(m_0,m_0;\,0) =  \partial_{x_\mathrm{C}} g(m_0,m_0;\,0)$, so $a_2=1/8$. At last, if the rates are constants, also $g(x_\mathrm{H},x_\mathrm{C};\,\eta_\mathrm{c})$ is constant, trivially satisfying $\partial_{x_\mathrm{H}} g =  \partial_{x_\mathrm{C}} g = 0$.

\section*{Appendix D. COP at Maximum Power}
 \setcounter{section}{4}
 \setcounter{equation}{0}
 \label{app_d}
In this appendix we prove Eqs.~(9) and (10) of the main text  and we derive the scaling of the COP at maximum power for large values of the maximum gap $\Delta$ given by $\mathrm{C}_{\mathrm{op}}(P^{\max}_{[\mathrm{R}]}) \propto 1/(\beta_\mathrm{C}\Delta)$. The COP at maximum power can be written as [see Eq.~(8) of the main text]
\begin{eqnarray}
	\mathrm{C}_{\mathrm{op}}(P^{\max}_{[\mathrm{R}]}) = \frac{\epsilon^*_\mathrm{C}}{\epsilon^*_\mathrm{H}-\epsilon^*_\mathrm{C}},
	\label{eq:cop_eps}
\end{eqnarray}
where $\epsilon^*_\mathrm{H}$ and $\epsilon^*_\mathrm{C}$ are respectively the values of $\epsilon_\mathrm{H}$ and $\epsilon_\mathrm{C}$ that maximize [see Eq.~(5) of the main text]
\begin{equation}
	 P_{[\mathrm{R}]}(\epsilon_\mathrm{H},\epsilon_\mathrm{C}) := - g(\epsilon_\mathrm{H},\epsilon_\mathrm{C}) \epsilon_\mathrm{C} \left[f(\beta_\mathrm{H}\epsilon_\mathrm{H}) - f(\beta_\mathrm{C}\epsilon_\mathrm{C}) \right],
	\label{eq:pr_eps}
\end{equation}
where $f(x):=[1+\exp{(x)}]^{-1}$ and
\begin{equation}
	g(\epsilon_\mathrm{H},\epsilon_\mathrm{C}) :=  \frac{\Gamma_\mathrm{H}(\epsilon_\mathrm{H})\Gamma_\mathrm{C}(\epsilon_\mathrm{C})}{ \left(\sqrt{\Gamma_\mathrm{H}(\epsilon_\mathrm{H})} +\sqrt{\Gamma_\mathrm{C}(\epsilon_\mathrm{C})} \right) ^2}.
\end{equation}

We first prove that the COP at maximum power takes the universal form of Eq.~(9) of the main text if the rates depend on the energy and on the temperature only through $\beta\epsilon$, i.e. $\Gamma_\alpha(\epsilon) = \Gamma_\alpha(\beta_\alpha \epsilon_\alpha)$. We rewrite Eq.~(\ref{eq:cop_eps}) as a function of  $x^*_\alpha = \beta_\alpha\epsilon^*_\alpha$ (for $\alpha = $ H, C):
 \begin{eqnarray} 
	\mathrm{C}_{\mathrm{op}}(P^{\max}_{[\mathrm{R}]}) =\left[ \frac{x^*_\mathrm{H}}{x^*_\mathrm{C}}\left(\frac{1}{\mathrm{C}_{\mathrm{op}}^{(\mathrm{c})}} + 1 \right) -1 \right]^{-1},
	\label{eq:cop_x}
\end{eqnarray} 
where $\mathrm{C}_{\mathrm{op}}^{(\mathrm{c})}$ is the Carnot COP for a refrigerator (see main text). We can determine $x^*_\alpha$ by maximizing
\begin{equation}
	 P_{[\mathrm{R}]}(x_\mathrm{H},x_\mathrm{C}) := -\frac{1}{\beta_\mathrm{C}}  \frac{\Gamma_\mathrm{H}(x_\mathrm{H})\Gamma_\mathrm{C}(x_\mathrm{C})}{ \left(\sqrt{\Gamma_\mathrm{H}(x_\mathrm{H})} +\sqrt{\Gamma_\mathrm{C}(x_\mathrm{C})} \right) ^2}  x_\mathrm{C} \left[ f(x_\mathrm{H}) - f(x_\mathrm{C}) \right].
	\label{eq:pr_x}
\end{equation}
Crucially, given our hypothesis on the rates, there is no explicit dependence on the temperatures in Eq.~(\ref{eq:pr_x}) (except for the prefactor $1/\beta_\mathrm{C}$), so the maximization of $P_{[\mathrm{R}]}(x_\mathrm{H},x_\mathrm{C})$ will simply yield two values of $x^*_\mathrm{H}$ and $x^*_\mathrm{C}$ that do not depend on the temperatures. Thus, for all bath temperatures the COP at maximum power will be given by Eq.~(\ref{eq:cop_x}), where $x^*_\mathrm{H}$ and $x^*_\mathrm{C}$ are two fixed values. The ratio $x^*_\mathrm{H}/x^*_\mathrm{C}$ will depend on the specific rates we consider. By imposing in Eq.~(\ref{eq:cop_x}) that the COP at maximum power of the system for $\beta_\mathrm{H}=\beta_\mathrm{C}$ (i.e. for $\mathrm{C}_{\mathrm{op}}^{(\mathrm{c})} \to \infty$) is $\mathrm{C}_{\mathrm{op}}^{(0)}$, we can eliminate the ratio $x^*_\mathrm{H}/x^*_\mathrm{C}$ in favor of $\mathrm{C}_{\mathrm{op}}^{(0)}$, concluding the proof of Eq.~(9) of the main text.

We now prove Eq.~(10) of the main text.  Since Eq.~(\ref{eq:pr_eps}) remains unchanged by sending both $\epsilon_\mathrm{H} \to -\epsilon_\mathrm{H}$ and $\epsilon_\mathrm{C} \to -\epsilon_\mathrm{C}$, we can assume without loss of generality that $\epsilon_\mathrm{C}\geq 0$ (this is a general property which 
applies independently of the specific choice of bath models). Furthermore, we must ensure that the system is acting as a refrigerator by imposing $ P_{[\mathrm{R}]}(\epsilon_\mathrm{H},\epsilon_\mathrm{C}) \geq 0$. This implies that $f(\beta_\mathrm{H}\epsilon_\mathrm{H})\leq f(\beta_\mathrm{C}\epsilon_\mathrm{C})$, thus
\begin{equation}
	0\leq \beta_\mathrm{C}\epsilon_\mathrm{C} \leq \beta_\mathrm{H}\epsilon_\mathrm{H}.
	\label{eq:ref_cond}
\end{equation}
We now show that in the models described by Eq.~(6) of the main text, the partial derivative of $ P_{[\mathrm{R}]}(\epsilon_\mathrm{H},\epsilon_\mathrm{C})$ respect to $\epsilon_\mathrm{H}$ is non negative for all values of $\epsilon_\mathrm{H}$ and $\epsilon_\mathrm{C}$ satisfying Eq.~(\ref{eq:ref_cond}), which implies that $\epsilon^*_\mathrm{H} \to +\infty$. Using Eq.~(\ref{eq:ref_cond}), the condition  $\partial P_{[\mathrm{R}]}(\epsilon_\mathrm{H},\epsilon_\mathrm{C})/\partial \epsilon_\mathrm{H} \geq 0$ can be written as
\begin{equation}
	\frac{\partial }{\partial \epsilon_\mathrm{H}} \ln{g(\epsilon_\mathrm{H}, \epsilon_\mathrm{C})} \geq -\frac{\beta_\mathrm{H}}{2\left[1+\cosh{(\beta_\mathrm{H}\epsilon_\mathrm{H})} \right]}.
	\label{eq:dp_deh}
\end{equation}
Since $\partial \ln{g(\epsilon_\mathrm{H}, \epsilon_\mathrm{C})}/ \partial \epsilon_\mathrm{H} $ has the same sign as $d \Gamma_\mathrm{H}(\epsilon_\mathrm{H})/ d \epsilon_\mathrm{H}$, and since the r.h.s. of Eq.~(\ref{eq:dp_deh}) is strictly negative, Eq.~(\ref{eq:dp_deh}) is certainly satisfied whenever $\Gamma_\mathrm{H}(\epsilon_\mathrm{H})$ is a growing function. This proves that $\epsilon^*_\mathrm{H} \to +\infty$ when the baths are described by the $\mathrm{F}_n$ model [see Eq.~(6) of the main text] even when the two baths have different powers $n$. The $\mathrm{B}_n$ model is more tricky to analyze since the rates are decreasing functions around the origin. Nonetheless, using Eq.~(\ref{eq:ref_cond}) it is possible to show that Eq.~(\ref{eq:dp_deh}) is satisfied also in the $\mathrm{B}_n$ model by plugging $\Gamma^{(\mathrm{B}_n)}_\alpha(\epsilon)$ [see Eq.~(6) of the main text] into Eq.~(\ref{eq:dp_deh}). This result holds also when the two baths have different powers $n$. 

We now know that $\epsilon^*_\mathrm{H} \to +\infty$ in the $\mathrm{F}_n$ and $\mathrm{B}_n$ models. Since both $\Gamma^{(\mathrm{F}_n)}_\mathrm{H}(\epsilon)$ and $\Gamma^{(\mathrm{B}_n)}_\mathrm{H}(\epsilon)$ diverge for $n>0$ when  $\epsilon^*_\mathrm{H} \to +\infty$, we have that
\begin{equation}
	g(+\infty,\epsilon_\mathrm{C}) =  \Gamma_\mathrm{C}(\epsilon_\mathrm{C}) = k_\mathrm{C}\, \epsilon_\mathrm{C}^n\, h(\beta_\mathrm{C}\epsilon_\mathrm{C}) =  k_\mathrm{C}\, \frac{x_\mathrm{C}^n}{\beta_\mathrm{C}^n} \, h(x_\mathrm{C}),
\end{equation}
where, as before, $x_\mathrm{C} = \beta_\mathrm{C}\epsilon_\mathrm{C}$ and $h(x):= 1$ for the $\mathrm{F}_n$ model and $h(x):= \coth{x/2}$ for the $\mathrm{B}_n$ model [see Eq.~(6) of the main text]. Thus, using $x_\mathrm{C}$ instead of $\epsilon_\mathrm{C}$, and noting that $f(\epsilon_\mathrm{H}\beta_\mathrm{H})$ vanishes for $\epsilon_\mathrm{H}\to +\infty$, we can write $P_{[\mathrm{R}]}(+\infty,\epsilon_\mathrm{C})$ [see Eq.~(\ref{eq:pr_eps})] as 
\begin{equation}
	 P_{[\mathrm{R}]}^{(n>0)} =   \frac{k_\mathrm{C}}{\beta_\mathrm{C}^{n+1}} \, x_\mathrm{C}^{n+1} h(x_\mathrm{C}) f(x_\mathrm{C}).
	 \label{eq:prmax_ninf}
\end{equation}
Equation~(\ref{eq:prmax_ninf}) is non-negarive for all values of $x_\mathrm{C}$ and it vanishes in $x_\mathrm{C} = 0$ and   $x_\mathrm{C} \to +\infty$ thanks to the exponential decrease of $f(x_\mathrm{C})$ for large values of $x_\mathrm{C}$. Therefore, Eq.~(\ref{eq:prmax_ninf}) will be maximum for the finite value $x^*_\mathrm{C}$ that maximizes $ x_\mathrm{C}^{n+1} h(x_\mathrm{C}) f(x_\mathrm{C})$, and plugging $x^*_\mathrm{C}$ into Eq.~(\ref{eq:prmax_ninf}) yields the first relation in Eq.~(10) of the main text, where $c_n = (x^*_\mathrm{C})^{n+1}h(x_\mathrm{C}^*) f(x^*_\mathrm{C})$. 
For $n=0$, we separately analyze the $\mathrm{F}_0$ and $\mathrm{B}_0$ models. In the $\mathrm{F}_0$ model, $g(\epsilon_\mathrm{H},\epsilon_\mathrm{C}) = k_\mathrm{H}k_\mathrm{C}/(\sqrt{k_\mathrm{H}} + \sqrt{k_\mathrm{C}})^2$, so $P_{[\mathrm{R}]}(+\infty,\epsilon_\mathrm{C})$ can be written as 
\begin{equation}
	 P_{[\mathrm{R}]}^{(\mathrm{F}_0)} =  \frac{k_\mathrm{C}}{\beta_\mathrm{C}}  \frac{r}{(\sqrt{r} + 1)^2} \, x_\mathrm{C} f(x_\mathrm{C}),
	 \label{eq:prmax_f0}
\end{equation}
where $r:= k_\mathrm{H}/k_\mathrm{C}$. Using the same argument as before, Eq.~(\ref{eq:prmax_f0}) implies a finite value of $x^*_\mathrm{C}$ which arises from the maximization of  $ x_\mathrm{C} f(x_\mathrm{C})$. We thus proved the first relation in Eq.~(10) of the main text for the $\mathrm{F}_0$ model, where $c_0 = r/(\sqrt{r} + 1)^2 \, x^*_\mathrm{C} f(x^*_\mathrm{C})$. At last, in the  $\mathrm{B}_0$ model  $g(\epsilon_\mathrm{H}\to +\infty,\epsilon_\mathrm{C}) = k_\mathrm{H}k_\mathrm{C}\coth{(x_\mathrm{C}/2)}/[\sqrt{k_\mathrm{H}} + \sqrt{k_\mathrm{C}\coth{(x_\mathrm{C}/2)}}]^2$. Thus, $P_{[\mathrm{R}]}( +\infty,\epsilon_\mathrm{C})$ can be written as
\begin{equation}
	 P_{[\mathrm{R}]}^{(\mathrm{B}_0)} =  \frac{k_\mathrm{C}}{\beta_\mathrm{C}}  \frac{r\coth{(x_\mathrm{C}/2)}}{(\sqrt{r} + \sqrt{\coth{(x_\mathrm{C}/2)}})^2} \, x_\mathrm{C} f(x_\mathrm{C}).
	 \label{eq:prmax_b0}
\end{equation}
Again, $x^*_\mathrm{C}$ is a finite value which can be found by maximizing $r\coth{(x_\mathrm{C}/2)}/(\sqrt{r} + \sqrt{\coth{(x_\mathrm{C}/2)}})^2 \, x_\mathrm{C} f(x_\mathrm{C})$. Only in this case, $x^*_\mathrm{C}$ depends on the ratio $r$. We thus proved the first relation in Eq.~(10) of the main text for the $\mathrm{B}_0$ model, where $c_0 = r\coth{(x^*_\mathrm{C}/2)}/(\sqrt{r} + \sqrt{\coth{(x^*_\mathrm{C}/2)}})^2 \, x^*_\mathrm{C} f(x^*_\mathrm{C})$.

The second relation in Eq.~(10) of the main text stems from the fact that in all models $\epsilon_\mathrm{H}^* \to +\infty$ while $\epsilon_\mathrm{C}^*$ is finite. Thus, Eq.~(\ref{eq:cop_eps}) implies that the $\mathrm{C}_{\mathrm{op}}(P^{\max}_{[\mathrm{R}]})$ vanishes.
At last we want to roughly estimate the behavior of $\mathrm{C}_{\mathrm{op}}(P^{\max}_{[\mathrm{R}]})$ in the presence of a large yet finite constraint on the maximum gap: $|\epsilon(t)|\leq \Delta$. Since $\epsilon_\mathrm{H}$ would diverge if there was no constraint, we can assume that, in the presence of $\Delta$, $\epsilon_\mathrm{H}^* = \Delta$. On the other hand, $\epsilon^*_\mathrm{C}$ is a finite quantity (which is given by $\epsilon^*_\mathrm{C} = x^*_\mathrm{C}/\beta_\mathrm{C}$ in the unconstrained case), so if we assume that $\Delta \gg \epsilon^*_\mathrm{C}$, from Eq.~(\ref{eq:cop_eps}) we have that
\begin{equation}
	\mathrm{C}_{\mathrm{op}}(P^{\max}_{[\mathrm{R}]}) \approx \frac{\epsilon^*_\mathrm{C}}{\epsilon^*_\mathrm{H}} \approx \frac{x^*_\mathrm{C}}{\beta_\mathrm{C}\Delta} \propto \frac{1}{\beta_\mathrm{C}\Delta}.
\end{equation}


\section*{References}

\begin{thebibliography}{1}
\bibitem{Kosloff2013}
Kosloff, R. Quantum thermodynamics: a dynamical viewpoint.
\href{https://doi.org/10.3390/e15062100}{\textit{Entropy {\bf 15}}, 2100-2128 (2013).}
\bibitem{Kosloff2014}
Kosloff, R. \& Levy, A. Quantum heat engines and refrigerators: continuous devices.
\href{https://doi.org/10.1146/annurev-physchem-040513-103724}{\textit{Annu. Rev. Phys. Chem.} {\bf 65}, 365-393 (2014).}
\bibitem{Benenti2017}
Benenti, G., Casati, G., Saito, K. \& Whitney, R. S. Fundamental aspects of steady-state conversion of heat to work at the nanoscale.
\href{https://doi.org/10.1016/j.physrep.2017.05.008}{\textit{Phys. Rep.} {\bf 694}, 1-124 (2017).}
\bibitem{Alicki2018}
Alicki, R. \& Kosloff, R. Introduction to quantum thermodynamics: history and prospects.
\href{https://doi.org/10.1007/978-3-319-99046-0_1}{\textit{Thermodynamics in the Quantum
Regime: Fundamental Aspects and New Directions} (Berlin, Springer, 2019).}
\bibitem{Buffoni2018}
Buffoni, L., Solfanelli, A., Verrucchi, P., Cuccoli, A. \& Campisi, M. Quantum Measurement Cooling.
\href{https://doi.org/10.1103/PhysRevLett.122.070603}{\textit{Phys. Rev. Lett.} {\bf 122}, 070603 (2019).}
\bibitem{Pekola2015} 
Pekola, J. P. Towards quantum thermodynamics in electronic circuits.
\href{https://doi.org/10.1038/nphys3169}{\textit{Nat. Phys.} {\bf 11}, 118-123 (2015).}
\bibitem{Goold2016}
Goold, J., Huber, M., Riera, A., del Rio, L. \& Skrzypczyk, P. The role of quantum information in thermodynamics-a topical review.
\href{https://doi.org/10.1088/1751-8113/49/14/143001}{\textit{J. Phys. A: Math. Theor.} {\bf 49}, 143001 (2016).}
\bibitem{Vinjanampathy2016}
Vinjanampathy, S. \& Anders, J. Quantum thermodynamics.
\href{https://doi.org/10.1080/00107514.2016.1201896}{\textit{Contemp. Phys.} {\bf 57}, 545-579 (2016).}
\bibitem{Chen1994}
 Chen, J. The maximum power output and maximum efficiency of an irreversible Carnot heat engine.
\href{https://doi.org/10.1088/0022-3727/27/6/011}{ \textit{J. Phys. D} {\bf 27}, 1144 (1994).}
\bibitem{Rezek2006} Rezek, Y. \& Kosloff, R. Irreversible performance of a quantum harmonic heat engine.
\href{https://doi.org/10.1088/1367-2630/8/5/083}{\textit{New J. Phys.} {\bf 8}, 83 (2006).}
\bibitem{Scully2011}
Scully, M. O., Chapin, K. R., Dorfman, K. E., Kim, M. B. \& Svidzinsky, A. Quantum heat engine power can be increased by noise-induced coherence.
\href{https://doi.org/10.1073/pnas.1110234108}{\textit{Proc. Natl. Acad. Sci. U.S.A.} {\bf 108}, 15097-15100 (2011).}
\bibitem{Abah2012}
Abah, O. et al. Single-ion heat engine at maximum power.
\href{https://doi.org/10.1103/PhysRevLett.109.203006}{\textit{Phys. Rev. Lett.} {\bf 109}, 203006 (2012).}
\bibitem{Correa2013}
Correa, L. A., Palao, J. P., Adesso, G. \& Alonso, D. Performance bound for quantum absorption refrigerators.
\href{https://doi.org/10.1103/PhysRevE.87.042131}{\textit{Phys. Rev. E} {\bf 87}, 042131 (2013).}
\bibitem{Dorfman2013}
Dorfman, K. E., Voronine, D. V., Mukamel, S. \& Scully, M. O. Photosynthetic reaction center as a quantum heat engine.
\href{https://doi.org/10.1073/pnas.1212666110}{\textit{Proc. Natl. Acad. Sci. U.S.A.} {\bf 110}, 2746-2751 (2013).}
\bibitem{Brunner2014} Brunner, N. et al. Entanglement enhances cooling in microscopic quantum refrigerators.
\href{https://doi.org/10.1103/PhysRevE.89.032115}{\textit{Phys. Rev. E} {\bf 89}, 032115 (2014).}
\bibitem{Zhang2014}
Zhang, K., Bariani, F. \& Meystre, P. Quantum optomechanical heat engine.
\href{https://doi.org/10.1103/PhysRevLett.112.150602}{\textit{Phys. Rev. Lett.} {\bf 112}, 150602 (2014).}
\bibitem{Campisi2016}
Campisi, M. \& Fazio, R. The power of a critical heat engine.
\href{https://doi.org/10.1038/ncomms11895}{\textit{Nat. Commun.} {\bf 7}, 11895 (2016).}
\bibitem{Rossnagel2016}
Ro{\ss}nagel, J. et al. A single-atom heat engine.
\href{https://doi.org/10.1126/science.aad6320}{\textit{Science} {\bf 352}, 325-329 (2016).}
\bibitem{Brandner2017}
Brandner, K., Bauer, M. \& Seifert U. Universal coherence-induced power losses of quantum heat engines in linear response.
\href{https://doi.org/10.1103/PhysRevLett.119.170602}{\textit{Phys. Rev. Lett.} {\bf 119}, 170602 (2017).}
\bibitem{Watanabe2017}
Watanabe, G., Venkatesh, B. P., Talkner, P. \& del Campo, A. Quantum performance of thermal machines over many cycles.
\href{https://doi.org/10.1103/PhysRevLett.118.050601}{\textit{Phys. Rev. Lett.} {\bf 118}, 050601 (2017).}
\bibitem{Josefsson2018}
Josefsson, M. et al. A quantum-dot heat engine operating close to the thermodynamic efficiency limits.
\href{https://doi.org/10.1038/s41565-018-0200-5}{\textit{Nat. Nanotechnol.} {\bf 13}, 920-924 (2018).}
\bibitem{Ronzani2018}
Ronzani, A. et al. Tunable photonic heat transport in a quantum heat valve.
\href{https://doi.org/10.1038/s41567-018-0199-4}{\textit{Nat. Phys.} {\bf 14}, 991-995 (2018).}
\bibitem{Prete2019}
Prete, D. et al. Thermoelectric conversion at 30 K in InAs/InP nanowire quantum dots.
\href{https://doi.org/10.1021/acs.nanolett.9b00276}{\textit{Nano Lett.} {\bf 19}, 3033-3039 (2019).}
\bibitem{Esposito2010bis}
Esposito, M., Kawai, R., Lindenberg, K. \&  Van den Broeck, C. Efficiency at maximum power of low-dissipation Carnot engines.
\href{https://doi.org/10.1103/PhysRevLett.105.150603}{\textit{Phys. Rev. Lett.} {\bf105}, 150603 (2010).}
\bibitem{Wang2011}
Wang, J., He, J. \& He, X. Performance analysis of a two-state quantum heat engine working with a single-mode radiation field in a cavity.
\href{https://doi.org/10.1103/PhysRevE.84.041127}{\textit{Phys. Rev. E} {\bf 84}, 041127 (2011).}
\bibitem{Ludovico2016}
Ludovico, M. F., Battista, F., von Oppen, F. \& Arrachea, L. Adiabatic response and quantum thermoelectrics for ac-driven quantum systems.
\href{https://doi.org/10.1103/PhysRevB.93.075136}{\textit{Phys. Rev. B} {\bf 93}, 075136 (2016).}
\bibitem{Cavina2017}
Cavina, V., Mari, A. \& Giovannetti, V. Slow dynamics and thermodynamics of open quantum systems.
\href{https://doi.org/10.1103/PhysRevLett.119.050601}{\textit{Phys. Rev. Lett.} {\bf 119}, 050601 (2017).}
\bibitem{Abiuso2018}
Abiuso, P. \& Giovannetti, V. Non-Markov enhancement of maximum power for quantum thermal machines.
\href{https://doi.org/10.1103/PhysRevA.99.052106}{\textit{Phys. Rev. A} {\bf 99}, 052106 (2019).}
\bibitem{Deng2013}
Deng, J., Wang, Q.-h., Liu, Z., H{\"a}nggi, P. \& Gong J. Boosting work characteristics and overall heat-engine performance via shortcuts to adiabaticity: Quantum and classical systems.
\href{https://doi.org/10.1103/PhysRevE.88.062122}{\textit{Phys. Rev. E} {\bf 88}, 062122 (2013).}
\bibitem{Torrontegui2013}
Torrontegui, E. et al. Chapter 2 - shortcuts to adiabaticity.
\href{https://doi.org/10.1016/B978-0-12-408090-4.00002-5}{\textit{Adv. At. Mol. Opt. Phys.} {\bf 62}, 117-169 (2013).}
\bibitem{Campo2014}
del Campo, A., Goold,  J. \& Paternostro, M. More bang for your buck: Super-adiabatic quantum engines.
\href{https://doi.org/10.1038/srep06208}{\textit{Sci. Rep.} {\bf 4}, 6208 (2014).}
\bibitem{Cakmak2018}
\c{C}akmak, B. \& M\"{u}stecapl{\i}o\u{g}lu, \"{O}. E. Spin quantum heat engines with shortcuts to adiabaticity.
\href{https://doi.org/10.1103/PhysRevE.99.032108}{\textit{Phys. Rev. E} {\bf 99}, 032108 (2019).}
\bibitem{Andresen1982} 
Rubin, M. H. \& Andresen, B. Optimal staging of endoreversible heat engines. 
\href{https://doi.org/10.1063/1.331592}{\textit{J. Appl. Phys.} {\bf 53}, 1 (1982).}
\bibitem{Song2006}
Song, H., Chen, L. \& Sun, F. Endoreversible heat-engines for maximum power-output with fixed duration and radiative heat-transfer law.
\href{https://doi.org/10.1016/j.apenergy.2006.09.003}{\textit{Appl. Energy} {\bf 84}, 374-388 (2007).}
\bibitem{Mukherjee2013}
Mukherjee, V. et al. Speeding up and slowing down the relaxation of a qubit by optimal control.
\href{https://doi.org/10.1103/PhysRevA.88.062326}{\textit{Phys. Rev. A} {\bf 88}, 062326 (2013).}
\bibitem{Bonnard2009}
Bonnard, B., Chyba, M. \& Sugny, D. Time-minimal control of dissipative two-level quantum systems: the generic case.
\href{https://doi.org/10.1109/TAC.2009.2031212}{\textit{IEEE Trans. Aut. Control} {\bf 54}, 2598-2610 (2009).}
\bibitem{Zhang2015}
Zhang, T. M., Wu, R. B., Zhang, F. H., Tarn, T. J. \& Long, G. L. Minimum-time selective control of homonuclear spins.
\href{https://doi.org/10.1109/TCST.2015.2390191}{\textit{IEEE Trans. Control Syst. Technol.} {\bf 23}, 2018-2025 (2015).}
\bibitem{Roloff2009}
Roloff, R., Wenin, M. \& P\"otz, W. Optimal control for open quantum systems: qubits and quantum gates.
\href{https://doi.org/10.1166/jctn.2009.1246}{\textit{J. Comp. Theor. Nanoscience} {\bf 6}, 1837-1863 (2009).}
\bibitem{Schulte2011}
Schulte-Herbr\"uggen, T., Sp{\" o}rl, A., Khaneja, N. \& Glaser, S. J. Optimal control for generating quantum gates in open dissipative systems.
\href{https://doi.org/10.1088/0953-4075/44/15/154013}{\textit{J. Phys. B: At. Mol. Opt. Phys.} {\bf 44}, 154013 (2011).}
\bibitem{Carlini2006}
Carlini, A., Hosoya, A., Koike, T. \& Okudaira, Y. Time-optimal quantum evolution.
\href{https://doi.org/10.1103/PhysRevLett.96.060503}{\textit{Phys. Rev. Lett.} {\bf 96}, 060503 (2006).}
\bibitem{Sauer2013}
Sauer, S., Gneiting, C. \& Buchleitner, A. Optimal coherent control to counteract dissipation.
\href{https://doi.org/10.1103/PhysRevLett.111.030405}{\textit{Phys. Rev. Lett.} {\bf 111}, 030405 (2013).}
\bibitem{Campisi2015}
Campisi, M., Pekola, J. \& Fazio, R. Nonequilibrium fluctuations in quantum heat engines: theory, example, and possible solid state experiments.
\href{https://doi.org/10.1088/1367-2630/17/3/035012}{\textit{New J. Phys.} {\bf 17}, 035012 (2015).}
\bibitem{Cavina2017bis}
Cavina, V., Mari, A., Carlini, A. \& Giovannetti, V. Variational approach to the optimal control of coherently driven, open quantum system dynamics.
\href{https://doi.org/10.1103/PhysRevA.98.052125}{\textit{Phys. Rev. A} {\bf 98}, 052125 (2018).}
\bibitem{Suri2017}
Suri, N., Binder, F. C., Muralidharan, B. \& Vinjanampathy, S. Speeding up thermalisation via open quantum system variational optimisation.
\href{https://doi.org/10.1140/epjst/e2018-00125-6}{\textit{Eur. Phys. J. Spec. Top.}  {\bf 227}, 203 (2018).}
\bibitem{Cavina2018}  
Cavina, V., Mari, A., Carlini, A. \& Giovannetti, V. Optimal thermodynamic control in open quantum systems.
\href{https://doi.org/10.1103/PhysRevA.98.012139}{\textit{Phys. Rev. A} {\bf 98}, 012139 (2018).}
\bibitem{Pekola2019}
Pekola, J. P., Karimi, B., Thomas, G. \& Averin, D. V. Supremacy of incoherent sudden cycles.
\href{https://doi.org/10.1103/PhysRevB.100.085405}{\textit{Phys. Rev. B} {\bf 100}, 085405 (2019).}
\bibitem{Menczel2019}
Menczel, P., Pyh{\"a}ranta, T., Flindt, C. \& Brandner, K. Two-stroke optimization scheme for mesoscopic refrigerators.
\href{https://doi.org/10.1103/PhysRevB.99.224306}{\textit{Phys. Rev. B} {\bf 99}, 224306 (2019).}
\bibitem{Gorini1976}
Gorini, V., Kossakowski, A. \& Sudarshan, E. C. G. Completely positive dynamical semigroups of N-level systems.
\href{https://doi.org/10.1063/1.522979}{\textit{J. Math. Phys} {\bf 17}, 821-825 (1976).}
\bibitem{Lindblad1976}
Lindblad, G. On the generators of quantum dynamical semigroups.
\href{https://doi.org/10.1007/BF01608499}{\textit{Commun. Math. Phys.} {\bf 48}, 119-130 (1976).}
\bibitem{Feldmann1996}
Feldmann, T., Geva, E., Kosloff, R. \& Salamon, P. Heat engines in finite time governed by master equations.
\href{https://doi.org/10.1119/1.18197}{\textit{Am. J. Phys.} {\bf 64}, 485-492 (1996).}
\bibitem{Feldmann2000}
Feldmann, T. \& Kosloff, R. Performance of discrete heat engines and heat pumps in finite time.
\href{https://doi.org/10.1103/PhysRevE.61.4774}{\textit{Phys. Rev. E} {\bf 61}, 4774-4790 (2000).}
\bibitem{Cerino2016}
Cerino, L., Puglisi, A. \& Vulpiani, A. Linear and nonlinear thermodynamics of a kinetic heat engine with fast transformations.
\href{https://doi.org/10.1103/PhysRevE.93.042116}{\textit{Phys. Rev. E} {\bf 93}, 042116 (2016).}
\bibitem{Curzon1975}
Curzon, F. L. \& Ahlborn, B. Efficiency of a Carnot engine at maximum power output.
\href{https://doi.org/10.1119/1.10023}{\textit{Am. J. Phys.} {\bf43}, 22-24 (1975).}
\bibitem{Novikov1957}
Novikov, I. I. Efficiency of an atomic power generating installation.
\textit{J. Nucl. Energy II} {\bf7}, 125D128 (1958).
\bibitem{Chambadal1957}
Chambadal, P. Les centrales nucleares.
\textit{A. Colin} {\bf4}, 1 (1957).
\bibitem{Broeck2005}
Van den Broeck, C. Thermodynamic Efficiency at Maximum Power.
\href{https://doi.org/10.1103/PhysRevLett.95.190602}{\textit{Phys. Rev. Lett.} {\bf 95}, 190602 (2005).}
\bibitem{Schmiedl2007}
Schmiedl, T. \& Seifert, U. Efficiency at maximum power: An analytically solvable model for stochastic heat engines.
\href{https://doi.org/10.1209/0295-5075/81/20003}{\textit{EPL} {\bf81}, 20003 (2007).}
\bibitem{Koski2014}
Koski, J. V., Maisi, V. F., Pekola, J. P. \& Averin, D. V. Experimental realization of a Szilard engine with a single electron.
\href{https://doi.org/10.1073/pnas.1406966111}{\textit{Proc. Natl. Acad. Sci. U.S.A.} {\bf 111}, 13786-13789 (2014).}
\bibitem{Maillet2018}
Maillet, O. et al. Optimal probabilistic work extraction beyond the free energy difference with a single-electron device.
\href{https://doi.org/10.1103/PhysRevLett.122.150604}{\textit{Phys. Rev. Lett.} {\bf 122}, 150604 (2019).}
\bibitem{Breuer2002}
Breuer, H. P. \& Petruccione, F. The Theory of Open Quantum Systems.
\href{https://doi.org/10.1093/acprof:oso/9780199213900.001.0001}{(Oxford University Press, Oxford, 2002).}
\bibitem{Geva1992}
Geva, E. \& Kosloff, R. A quantum-mechanical heat engine operating in finite time. A model consisting of spin-1/2 systems as the working fluid.
\href{https://doi.org/10.1063/1.461951}{\textit{J. Chem. Phys.} {\bf 96}, 3054-3067 (1992).}
\bibitem{Alicki1979}
Alicki, R. The quantum open system as a model of the heat engine.
\href{https://doi.org/10.1088/0305-4470/12/5/007}{\textit{J. Phys. A: Math. Gen} {\bf 12}, L103 (1979).}
\bibitem{Senior2019}
Senior, J. et. al. Heat rectification via a superconducting artificial atom. 
Preprint at \href{https://arxiv.org/abs/1908.05574}{https://arxiv.org/abs/1908.05574} (2019).
\bibitem{Esposito2010}
Esposito, M., Kawai, R., Lindenberg, K. \& Van den Broeck, C. Quantum-dot Carnot engine at maximum power.
\href{https://doi.org/10.1103/PhysRevE.81.041106}{\textit{Phys. Rev.  E} {\bf 81}, 041106 (2010).}
\bibitem{Esposito2009}
Esposito, M., Lindenberg, K. \& Van den Broeck, C. Thermoelectric efficiency at maximum power in a quantum dot.
\href{https://doi.org/10.1209/0295-5075/85/60010}{\textit{EPL} {\bf 85}, 60010 (2009).}
\bibitem{note1} In principle, one can consider a broader family of controls including the possibility of rotating the Hamiltonian eigenvectors;
however there is evidence that such an additional freedom does not help in two-level systems \cite{Cavina2018,Pekola2019}.
\bibitem{dann2018}
Dann, R., Levy, A. \& Kosloff, R. Time-dependent Markovian quantum master equation.
\href{https://doi.org/10.1103/PhysRevA.98.052129}{\textit{Phys. Rev. A} {\bf 98}, 052129 (2018).}
\bibitem{NOTA} This particular scaling has been found also in the optimization of endoreversible Carnot Heat engines \cite{Chen1994}.
\bibitem{Quan2007}
Quan, H. T., Liu, Y., Sun, C. P. \& Nori, F. Quantum thermodynamic cycles and quantum heat engines.
\href{https://doi.org/10.1103/PhysRevE.76.031105}{\textit{Phys. Rev. E} {\bf 76}, 031105 (2007).}
\bibitem{Karimi2016}
Karimi, B. \& Pekola, J. P. Otto refrigerator based on a superconducting qubit: Classical and quantum performance.
\href{https://doi.org/10.1103/PhysRevB.94.184503}{\textit{Phys. Rev. B} {\bf 94}, 184503 (2016).}
\bibitem{Kosloff2017}
Kosloff, R. \& Rezek, Y. The quantum harmonic Otto cycle.
\href{https://doi.org/10.3390/e19040136}{\textit{Entropy} {\bf 19}, 136 (2017).}
\bibitem{Beenakker1991}
Beenakker, C. W. J. Theory of Coulomb-blockade oscillations in the conductance of a quantum dot.
\href{https://doi.org/10.1103/PhysRevB.44.1646}{\textit{Phys. Rev. B} {\bf 44}, 1646 (1991).}
\bibitem{Erdman2017}  
Erdman, P. A. et al. Thermoelectric properties of an interacting quantum dot based heat engine.
\href{https://doi.org/10.1103/PhysRevB.95.245432}{\textit{Phys. Rev. B} {\bf 95}, 245432 (2017).}
\bibitem{Wang2012pre}
Wang, J., Wu, Z. \& He, J. Quantum Otto engine of a two-level atom with single-mode fields.
\href{https://doi.org/10.1103/PhysRevE.85.041148}{\textit{Phys. Rev. E} {\bf 85}, 041148 (2012).}
\bibitem{Wegewijs1999} Wegewijs, M. R. \& Nazarov, Y. V. Resonant tunneling through linear arrays of quantum dots.
\href{https://doi.org/10.1103/PhysRevB.60.14318}{\textit{Phys. Rev. B} {\bf 60}, 14318-14327 (1999).}
\bibitem{VanHorne2018}
Van Horne, N. et al. Single atom energy-conversion device with a quantum load.
Preprint at \href{https://arxiv.org/abs/1812.01303}{https://arxiv.org/abs/1812.01303} (2018).
\bibitem{Chicone1999}
Chicone, C. Ordinary Differential Equations with Applications. 
\href{https://doi.org/10.1007/b97645}{(Springer-Verlag, New York 1999).}
\end{thebibliography}
\end{document}